\documentclass{article}

\usepackage{arxiv}

\usepackage[utf8]{inputenc}
\usepackage[T1]{fontenc}
\usepackage{hyperref}
\usepackage{url}
\usepackage{booktabs}
\usepackage{amsfonts}
\usepackage{nicefrac}
\usepackage{microtype}
\usepackage{relsize}
\usepackage{verbatim}
\usepackage{graphicx}
\usepackage{dcolumn}
\usepackage{bm}
\usepackage{float}
\usepackage{graphics}
\usepackage{color}
\usepackage{xcolor}
\usepackage{algorithm}
\usepackage{algpseudocode}
\usepackage{subcaption}
\usepackage{tikz}
\usetikzlibrary{arrows.meta,positioning,fit,calc,shapes.geometric,backgrounds}
\usepackage{amsmath,amssymb,amsfonts}

\usepackage{algorithm}
\usepackage{algpseudocode}
\usepackage{enumitem}
\setlist[itemize]{leftmargin=*}
\setlist[enumerate]{leftmargin=*}
\usepackage{array}
\newcolumntype{P}[1]{>{\centering\arraybackslash}p{#1}}
\usepackage{multirow}
\usepackage[capitalise]{cleveref}

\usepackage{cite}
\usepackage{tabularx}
\usepackage{threeparttable}
\usepackage{algorithm}
\usepackage{algpseudocode}
\usepackage{pifont}

\newcolumntype{Y}{>{\centering\arraybackslash}X}

\newcommand{\R}{\mathbb{R}}

\newcommand{\G}{\mathcal{G}}

\title{Quantum SEDONet: Spectrally-Embedded Quantum Deep Operator Networks for Partial Differential Equations}

\author{
  Muhammad Abid \\
  Department of Mechanical and Aerospace Engineering\\
  University of Tennessee, Knoxville\\
  Knoxville, TN 37996, USA\\
  \texttt{mabid@vols.utk.edu}
  \And
  Arth Sojitra \\
  Department of Mechanical and Aerospace Engineering\\
  University of Tennessee, Knoxville\\
  Knoxville, TN 37996, USA\\
  \texttt{asojitra@vols.utk.edu}
  \And
  Bipin Tiwari \\
  Department of Mechanical and Aerospace Engineering\\
  University of Tennessee, Knoxville\\
  Knoxville, TN 37996, USA\\
  \texttt{btiwari1@vols.utk.edu}
  \And
  Omer San \\
  Department of Mechanical and Aerospace Engineering\\
  University of Tennessee, Knoxville\\
  Knoxville, TN 37996, USA\\
  \texttt{osan@utk.edu}
}

\begin{document}
\maketitle

\begin{abstract}
Quantum DeepONet accelerates neural-operator inference by evaluating an
orthogonally parameterized network on a quantum computer, reproducing in ideal
simulation the accuracy of its classical counterpart at asymptotically lower
inference cost. Its trunk network, however, receives the query coordinate raw,
so the multilayer perceptron must synthesize all spectral structure through its
nonlinearities; because coordinate networks are biased toward low frequencies,
the approximation error concentrates precisely where the solution varies
fastest. We propose Quantum SEDONet (Spectral-Embedded Deep Operator Network),
which assigns each trunk coordinate the classical spectral basis its boundary
condition dictates: a Fourier expansion for periodic coordinates and a Chebyshev
expansion for bounded, non-periodic ones. Two properties distinguish this from
existing coordinate embeddings. First, the basis is selected per coordinate
rather than per problem, so a single problem may carry both, and the choice
follows from the boundary condition rather than from any special role of time.
Second, and specific to the quantum setting, the embedding is free in the
resource that constrains the platform: under unary amplitude encoding, a layer's
qubit count is one plus the maximum of its input and output dimensions,
and the network width already saturates that maximum, so an embedding at or
below the width incurs zero additional qubits and no change in circuit depth,
for a parameter increase of a few percent. On four benchmarks the method reduces the mean relative $L^2$ error by
$54.1\%$ (antiderivative), $49.6\%$ (advection), $36.0\%$ (Burgers), and
$36.2\%$ (a mixed-boundary channel Poisson problem), demonstrating consistent
improvements over the baseline, while the quantum and classical evaluation
paths agree to within $10^{-8}$ throughout. The channel Poisson case
exercises both bases at once, Fourier for the periodic coordinate and Chebyshev
for the walled one, confirming that the rule is driven by each coordinate's
boundary condition. A representational upgrade that would cost compute on
classical hardware is therefore obtained at no quantum-resource cost.
\end{abstract}

\textbf{Keywords:}
Scientific Machine Learning (SciML); Neural Operators; Quantum Computing; Deep Operator Networks; Spectral Methods; Partial Differential Equations

\section{Introduction}
\label{sec:intro}
 
The numerical solution of partial differential equations (PDEs) underlies
computational science and engineering across fluid dynamics, structural
mechanics, heat transfer, and reaction
chemistry~\cite{leveque2002finite,quarteroni2008numerical}.
Finite element, finite difference, finite volume, and spectral element
discretizations are accurate and
well-understood~\cite{quarteroni2008numerical,karniadakis2005spectral,fornberg1996practical},
but they are expensive in a specific and consequential way: the cost is incurred
per solve. Any change to the boundary data, the forcing, the material
parameters, or the geometry requires resolving the entire discretized system,
which makes real-time and many-query workflows
impractical~\cite{hesthaven2016certified}. The bottleneck is
most severe in parametric studies, uncertainty
quantification~\cite{xiu2002wiener}, design optimization,
and digital-twin applications~\cite{rasheed2020digital,san2021hybrid},
where thousands of PDE instances must be solved and where the marginal cost of
one more solve determines whether the workflow is feasible at all.
 
Reduced-order modeling has long been the standard response. Projection-based
methods construct a low-dimensional subspace from solution snapshots by proper
orthogonal decomposition~\cite{benner2015survey},
dynamic mode decomposition~\cite{schmid2010dynamic}, or operator
inference~\cite{peherstorfer2016data}, and integrate the governing equations
within it. Non-intrusive variants replace the projection step with a regression
from parameters to reduced
coordinates~\cite{hesthaven2018nonintrusive,sojitra2025mml}. These methods are
mature and, within their regime of validity, extremely effective. Their
limitations are equally well characterized: a linear trial subspace cannot
efficiently represent advection-dominated dynamics whose Kolmogorov $n$-width
decays slowly, and the reduced model inherits the training distribution, so
extrapolation is unreliable. Nonlinear manifold
approaches relax the first constraint at the cost of an
opaque parameterization~\cite{lee2020model}.
 
Neural operators offer a different trade: rather than reducing a fixed
discretization, they learn a mapping between infinite-dimensional function
spaces directly, and are therefore mesh-free and resolution-flexible at
inference~\cite{kovachki2023neural,lu2021learning,li2021fourier,lanthaler2022error}.
Once trained, a neural operator evaluates a PDE solution orders of magnitude
faster than a numerical solver, which is the property that motivates its use as
a surrogate. The Deep Operator Network
(DeepONet)~\cite{lu2021learning} realizes such a mapping through two
subnetworks: a branch network that encodes the input function sampled at
fixed sensor locations, and a trunk network that encodes the query
coordinate at which the output is evaluated. The prediction is the inner product
of the two subnetwork outputs, a structure justified by a
universal-approximation theorem for operators~\cite{chen1995universal,lu2021learning}
and refined by subsequent error
analysis~\cite{lanthaler2022error}. The Fourier Neural Operator
(FNO)~\cite{li2021fourier} takes a complementary route, parameterizing the
integral kernel in the frequency domain, and a large family of architectural
variants has followed~\cite{li2023fourier,wen2022u,cao2023lno,
raonic2023convolutional,hao2023gnot}.
Operator learning has since been deployed across weather and climate
modeling~\cite{pathak2022fourcastnet}, subsurface flow~\cite{wen2022u},
seismic inversion~\cite{zhu2023fourierdeeponet}, fracture
mechanics~\cite{goswami2022fracture}, and
astrophysics~\cite{mao2023ppdonet}, and has been extended to multiple inputs,
multiple fidelities, and physics-informed
training~\cite{jin2022mionet,lu2022multifidelity,wang2021learning,wang2022improved}.
 
A separate line of work asks whether quantum computers can accelerate this
inference step. The motivation is that the noisy intermediate-scale quantum
(NISQ) era~\cite{preskill2018quantum,bharti2022noisy} offers devices whose
qubit counts and coherence times are limited but growing, and whose most
plausible near-term applications are those requiring shallow circuits and
modest register
widths~\cite{kim2023evidence}. Quantum algorithms with proven
asymptotic advantage exist for linear systems~\cite{harrow2009quantum},
for linear and dissipative nonlinear differential
equations~\cite{childs2021high,liu2021efficient},
and for fluid problems in particular~\cite{gaitan2020finding}, though the
input--output and state-preparation overheads temper their practical
reach~\cite{aaronson2015read}. On the machine-learning side, quantum models have
been proposed for classification, kernel methods, and regression on
differential
equations~\cite{schuld2019quantum,havlicek2019supervised,mengoni2019kernel,
suzuki2020feature,paine2023quantum,lubasch2020variational,chen2026qpinn},
with expressivity governed in large part by how classical data is encoded into
the quantum
state~\cite{schuld2021effect,perezsalinas2020data,jaderberg2024let,xiong2025qelm}.
Trainability, however, is a serious issue: barren plateaus induced by
circuit depth, cost-function locality, entanglement, and hardware noise flatten
the optimization landscape
exponentially~\cite{mcclean2018barren,larocca2025barren}.
 
Quantum DeepONet~\cite{xiao2025quantum} sidesteps the trainability problem by a
deliberate division of labor. Each dense layer of the branch and trunk is
replaced by an orthogonal layer built from a pyramid of two-qubit
reconfigurable beam-splitter (RBS)
gates~\cite{kerenidis2022quantum,landman2022quantum,cherrat2024quantum};
because such a layer has an exact classical counterpart, the network is trained
classically and only the trained angles are transferred to hardware for
inference. An orthogonal layer applied to a unary-encoded amplitude vector is
evaluated by a circuit whose depth grows linearly in the layer width, giving an
asymptotic inference speed-up over classical matrix multiplication. Crucially,
in the ideal (noiseless, infinite-shot) regime, the quantum evaluation reproduces
the classical orthogonal network exactly, so the two share a single accuracy
figure: the quantum computer changes where the network is evaluated, not
what function it computes~\cite{xiao2025quantum}.
 
This work starts from an observation about the trunk network rather than the
quantum layer. In the released architectures the trunk receives its coordinate
input either completely raw, as in the advection example where the pair $(x,t)$
is passed directly, or with a raw temporal coordinate, as in the Burgers
example. In both cases at least one coordinate reaches the trunk with no
spectral structure, and the multilayer perceptron (MLP) must reconstruct every
oscillation of the solution in that coordinate from a linear ramp. Coordinate
MLPs are biased toward low
frequencies, a phenomenon documented theoretically and empirically under the
names spectral bias and the frequency
principle~\cite{rahaman2019spectral,xu2020frequency}, and explained through the neural
tangent kernel~\cite{jacot2018neural,wang2021eigenvector}.
Consequently, the error concentrates wherever the solution has fine structure. We
confirm this empirically: for the Burgers data, the mean absolute error at $t=0$
is roughly twenty times that in the interior.
 
Our contribution, Quantum SEDONet, gives each trunk coordinate the
spectral basis its boundary condition calls for. Periodic coordinates receive a
Fourier expansion; bounded, non-periodic coordinates receive a Chebyshev
expansion~\cite{trefethen2000spectral,boyd2001chebyshev},
whose nodes cluster near the interval endpoints exactly where the boundary-layer
error resides. This is the classical prescription of spectral methods, imported
into the trunk input map rather than into the discretization. The essential
point for the quantum setting is a cost argument: unary encoding makes a layer's
qubit count equal to one plus the larger of its input and output dimensions.
Because the network width already dominates the trunk input dimension, an
embedding of dimension up to the width is absorbed for free, no extra qubits,
no extra circuit depth, and a parameter increase of a few percent. The method is
therefore not a trade of accuracy against quantum resources; it is a strict
improvement in the regime that matters for near-term
hardware~\cite{preskill2018quantum,bharti2022noisy}.

In this work, we make the following contributions.
\begin{enumerate}
\item \textbf{A boundary-matched basis-selection rule for the trunk.} We assign
each query coordinate the classical spectral basis its boundary condition
dictates, a Fourier expansion for periodic coordinates and a Chebyshev expansion
for bounded non-periodic ones, applied per coordinate rather than per problem.
Existing practice supplies a spectral input map only where a coordinate is
periodic and leaves every other coordinate raw; stating the rule for both cases
turns an occasional device into a systematic prescription, and the channel
Poisson benchmark shows the rule is driven by the boundary condition rather than
by any special role of time.
\item \textbf{The embedding is free in the resource that constrains the
platform.} Under unary amplitude encoding a layer's qubit count is one plus the maximum of its input and output dimensions, and the network width already
saturates that maximum. An embedding whose dimension stays at or below the width
therefore costs no additional qubits and no additional circuit depth, for a
parameter increase of a few percent. On classical hardware the same upgrade
widens the first layer and costs compute proportionally; in this architecture it
is absorbed within a well-defined budget. Quantum SEDONet is therefore not a
trade of accuracy against quantum resources but a strict improvement in the
regime relevant to near-term hardware.
\item \textbf{Consistent gains across four benchmarks at identical qubit count.}
On the antiderivative operator, advection, Burgers, and a mixed-boundary channel
Poisson problem, the embedding reduces the mean relative $L^2$ error by between
$36\%$ and $54\%$, improving on the baseline for the overwhelming majority of
test functions in every case under a paired comparison, while the quantum and
classical evaluation paths agree to within the precision of the classical
arithmetic.
\end{enumerate}

The remainder of this paper is organized as follows.
Section~\ref{sec:related} reviews related work on neural operators and
reduced-order modeling, spectral bias and coordinate embeddings, and quantum
computing for differential equations and operator learning.
Section~\ref{sec:methods} presents the Quantum SEDONet architecture in full
detail, covering the reconfigurable beam-splitter gate, unary data loading and
the numerically stable loader angles, the pyramidal orthogonal circuit, the
tomography that returns a classical vector, the assembly of these layers into a
quantum orthogonal neural network and into the Quantum DeepONet architecture,
the boundary-matched spectral trunk embedding and its qubit accounting, and the
training procedure.
Section~\ref{sec:results} presents the experimental setup, results, and
discussion for the function-approximation tests and all four PDE benchmarks,
including prediction visualizations, quantitative comparisons, the qubit and
parameter accounting, and evidence that the mechanism is consistent across
problems, followed by a section-level summary.
Section~\ref{sec:summary} provides the overall summary and conclusions, and
Section~\ref{sec:future} outlines directions for future work.
 
\section{Related Work}
\label{sec:related}

\subsection{Neural operators and reduced-order modeling}
Classical discretizations, finite difference, finite volume, finite element,
and spectral or spectral-element
methods~\cite{leveque2002finite,quarteroni2008numerical,karniadakis2005spectral}, remain
the reference against which surrogates are measured and supply the ground truth
for every benchmark in this paper. Their per-solve cost motivated decades of work
on reduced-order models: proper orthogonal decomposition and its
variants~\cite{benner2015survey},
dynamic mode decomposition~\cite{schmid2010dynamic}, operator
inference~\cite{peherstorfer2016data}, certified reduced-basis
methods~\cite{hesthaven2016certified}, non-intrusive regression
onto reduced coordinates~\cite{hesthaven2018nonintrusive,sojitra2025mml}, and
autoencoder-based nonlinear manifolds~\cite{lee2020model}. Neural operators
differ from all of these in that the learned object is a map between function
spaces rather than a reduced representation of one discretization, which is what
makes resolution-flexible inference possible.

DeepONet established the branch--trunk architecture and inherited a
universal-approximation guarantee for
operators~\cite{lu2021learning,chen1995universal}, later supplemented by
quantitative error estimates and cost--accuracy
analyses~\cite{lanthaler2022error}. FNO parameterized the kernel
in Fourier space~\cite{li2021fourier}, and subsequent architectures extended the
idea to general geometries~\cite{li2023fourier}, multiphase
flow~\cite{wen2022u}, the Laplace domain~\cite{cao2023lno}, multiwavelet and
wavelet bases~\cite{abid2026wlno}, U-shaped and
convolutional formulations~\cite{raonic2023convolutional}, 
attention-based operator
learning~\cite{hao2023gnot} and other recent operator
learning~\cite{sojitra2026fedonet, abid2026simr, abid2025spectral}.
Applications now span weather forecasting~\cite{pathak2022fourcastnet},
subsurface and multiphase transport~\cite{wen2022u},
full-waveform inversion~\cite{zhu2023fourierdeeponet}, fracture
mechanics~\cite{goswami2022fracture}, multiphysics coupling~\cite{cai2021deepm},
and protoplanetary dynamics~\cite{mao2023ppdonet}. Our embedding is orthogonal to
the choice of operator backbone: it modifies only how the query coordinate enters
the trunk and could in principle be combined with any architecture that has an
explicit coordinate input.

Several works modify DeepONet's subnetworks directly. MIONet generalizes the
branch to multiple input functions via a tensor product~\cite{jin2022mionet};
multifidelity formulations combine coarse and fine
data~\cite{lu2022multifidelity}; physics-informed
training replaces or supplements labeled data with the PDE
residual~\cite{wang2021learning,raissi2019physics};
improved training schemes address the imbalance between branch and trunk
gradients~\cite{wang2022improved}; and SVD-based analyses interpret the trunk
output as a learned basis~\cite{venturi2023svd}. The last of these is the closest
in spirit to our work: if the trunk is understood as producing a basis in which
the branch supplies coefficients, then the trunk's input representation
determines which bases are reachable, and supplying a spectral input map is a
direct way to control that. We follow the fair-comparison methodology of holding all non-embedding
hyperparameters fixed across arms~\cite{lu2022comprehensive}.

\subsection{Spectral bias, coordinate embeddings, and spectral bases}
The tendency of neural networks to fit low-frequency components before
high-frequency ones is established both empirically and
theoretically~\cite{rahaman2019spectral,xu2020frequency},
and is explained by the decay of the neural tangent kernel
eigenspectrum~\cite{jacot2018neural,wang2021eigenvector}. In
coordinate-based networks the standard remedy is an input feature embedding:
random Fourier features and their learned variants map a low-dimensional
coordinate into a higher-dimensional periodic feature space, dramatically
accelerating the learning of high-frequency
detail~\cite{tancik2020fourier,sitzmann2020implicit}.
Our Fourier trunk terms are a deterministic, low-order instance of this idea.
Our Chebyshev terms extend it to non-periodic coordinates, where a Fourier basis
would impose a spurious periodicity on a solution that satisfies Dirichlet or
initial conditions, an error that manifests as Gibbs oscillations near the
boundary. No prior operator-learning work selects the basis per
coordinate according to its boundary condition~\cite{sojitra2026fedonet,abid2025spectral}.

Fourier series are the natural expansion for periodic problems, while Chebyshev
polynomials provide a natural global basis for bounded, non-periodic problems, where
Chebyshev polynomial features can efficiently represent boundary-sensitive and
endpoint-sensitive structure~\cite{trefethen2000spectral,boyd2001chebyshev,fornberg1996practical}.
What is transplanted here is not the discretization but the basis
selection rule: the same criterion that governs whether a spectral solver uses
Fourier or Chebyshev in a given direction governs which features the trunk
receives for that coordinate. The transplant is not automatic, because the two
settings impose different constraints. A spectral solver chooses its truncation
to resolve the solution and pays for extra modes in linear algebra
cost~\cite{karniadakis2005spectral}; a trunk
embedding is bounded above by the network width and, under unary encoding, pays
nothing until that bound is reached~\cite{kerenidis2022quantum,xiao2025quantum}.
The design question therefore becomes one
of allocating a fixed feature budget across coordinates rather than of
convergence rate, which is why the same basis that would be an obvious choice
for a solver is not an obvious choice for a trunk until the resource accounting
is done.

Encoding boundary information into the architecture rather than the loss has an
established history in scientific machine learning. Hard-constraint formulations
multiply the network output by a distance function that vanishes on the Dirichlet
boundary, so the condition is satisfied by construction rather than
penalized~\cite{lu2022comprehensive}, and periodic problems are routinely handled
by substituting a truncated Fourier basis for the raw
coordinate~\cite{tancik2020fourier,sojitra2026fedonet}. Our contribution is to state the rule for
both boundary types and apply it per coordinate: periodic directions receive a
Fourier expansion and bounded non-periodic ones a Chebyshev expansion, the latter
addressing a boundary-layer failure mode that a Fourier basis
cannot~\cite{boyd2001chebyshev}. Assigning
the basis coordinate-wise rather than problem-wise is what turns an occasional
device into a systematic prescription.

\subsection{Quantum computing for differential equations and operator learning}
Quantum algorithms with proven asymptotic advantage exist for linear
systems~\cite{harrow2009quantum} and, building on them, for linear PDEs and for
dissipative nonlinear differential
equations~\cite{childs2021high,liu2021efficient},
including fluid applications~\cite{gaitan2020finding} and variational
formulations for nonlinear problems~\cite{lubasch2020variational,chen2026qpinn}. Modern
constructions rest on block-encoding and quantum singular value
transformation~\cite{gilyen2019quantum}. These approaches
address the solution; the present work, like Quantum
DeepONet~\cite{xiao2025quantum} and the quantum Fourier neural
operator~\cite{jain2023quantum}, instead accelerates the inference of a
trained surrogate, which avoids the state-preparation and readout overheads that
limit end-to-end quantum PDE
solvers~\cite{aaronson2015read}.

Orthogonal and unitary neural layers have been studied classically for their
stable gradients and improved
conditioning~\cite{arjovsky2016unitary}. The quantum
realization used here builds an orthogonal transformation from a pyramid of RBS
gates acting on a unary-encoded state, following the quantum orthogonal neural
network line of
work~\cite{kerenidis2022quantum,landman2022quantum,cherrat2024quantum}.
Variational and kernel-based quantum models form a parallel
literature~\cite{schuld2019quantum,havlicek2019supervised,mengoni2019kernel,
suzuki2020feature,altares2024genetic,assouel2022qgan,tao2025pulse},
whose central difficulty is trainability: barren plateaus arising from depth,
cost locality, entanglement, and
noise~\cite{mcclean2018barren,larocca2025barren}.
Quantum DeepONet avoids this entirely by training classically and transferring
angles, and supplies the tomography procedure that reads a layer's output back
into a classical vector~\cite{xiao2025quantum}. Where readout
is performed by finite sampling, error mitigation becomes
relevant~\cite{cai2023quantum};
the post-selection on unary bitstrings used in ~\cite{xiao2025quantum} is a
particularly cheap instance enabled by the encoding. Data encoding also
determines the expressivity of the resulting
model~\cite{schuld2021effect,perezsalinas2020data,jaderberg2024let,xiong2025qelm}.

The literature on encoding asks which feature maps a parameterized circuit can
realize and how the choice of embedding shapes the accessible function
class~\cite{schuld2021effect,perezsalinas2020data,suzuki2020feature}, with the frequency content of
the encoding emerging as the controlling factor~\cite{jaderberg2024let,xiong2025qelm}. What
distinguishes the unary orthogonal setting is that the register width is set by
the maximum of a layer's input and output dimensions rather than by the
input alone, so enriching the input costs nothing until it exceeds the network
width. A design decision that on classical hardware trades accuracy against
compute is therefore free within a well-defined budget, and the comparison
isolates the embedding as the sole source of any difference in accuracy.

\section{Methods}
\label{sec:methods}
In this section we develop the quantum foundation on which Quantum SEDONet is built. We first describe the quantum circuit used for a single network layer (\cref{sec:qlayer}), covering the reconfigurable beam-splitter gate, the loading of classical data into a unary-encoded state, the pyramidal orthogonal transformation, and the tomography that returns the result to a classical vector. We then assemble these layers into a quantum orthogonal neural network (\cref{sec:qorthonn}) and into the Quantum DeepONet architecture (\cref{sec:qdeeponet}). Building on this, we introduce Quantum SEDONet (\cref{sec:qsedonet}) and its boundary-matched spectral trunk embedding, and finally describe how the model is trained (\cref{sec:training}). Throughout, all results are reported in the ideal (noiseless, infinite-shot)
regime, in which the quantum evaluation reproduces the classical orthogonal
network exactly.

\subsection{Quantum methods for network layers}
\label{sec:qlayer}
\begin{figure}[H]
\centering
\includegraphics[width=\textwidth]{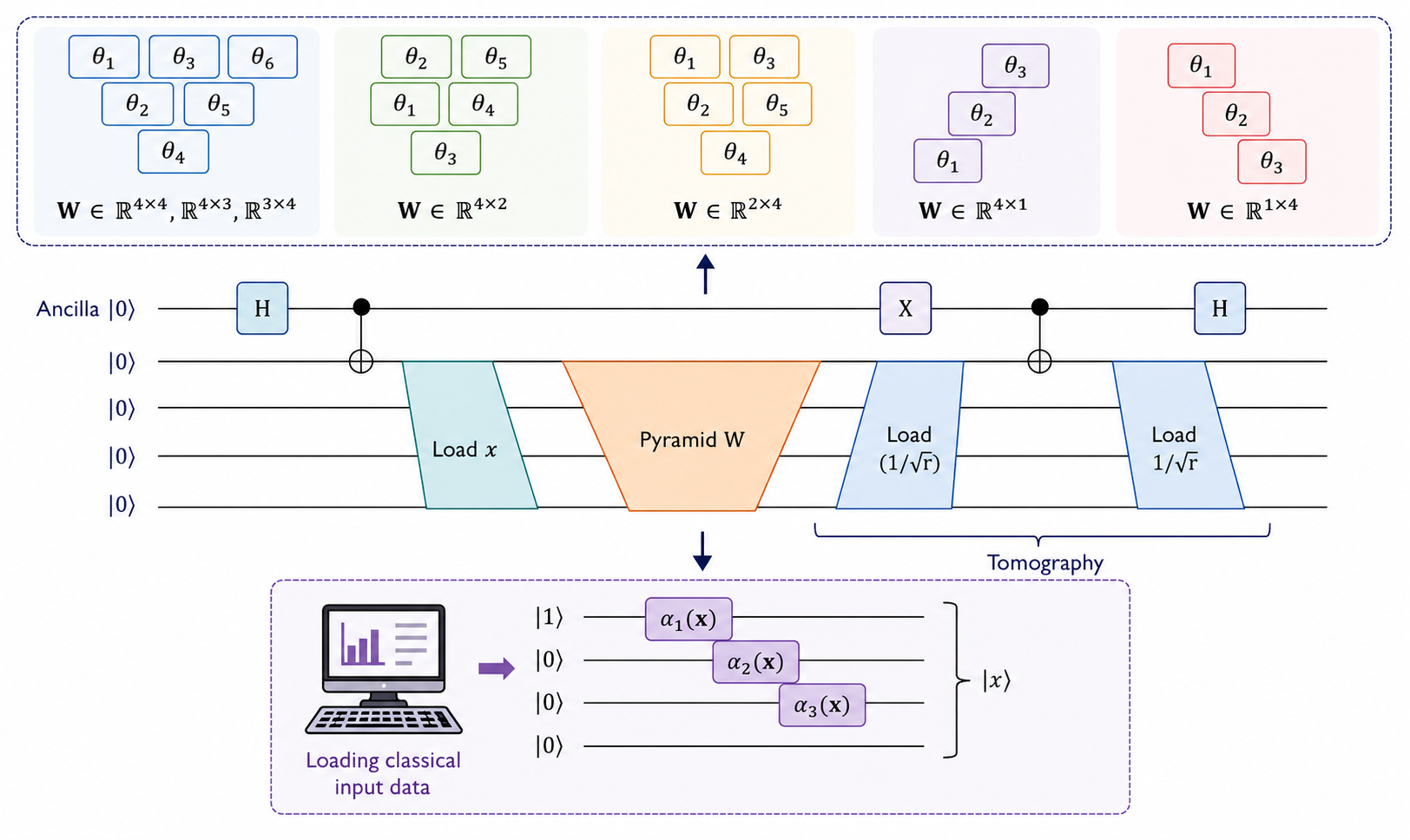}
\caption{The circuit of a single quantum layer, comprising data loading, a pyramidal orthogonal transformation, and tomography, with one ancillary qubit for the tomography. Vertical connectors denote the two-qubit RBS gates; $\theta_i$ parameterize the pyramid and $\alpha_i$ the loader. The top panels show the pyramid geometry for representative weight shapes $\mathbf{W}\in\R^{m\times n}$.}
\label{fig:quantum_layer}
\end{figure}

A classical neural-network layer with input $\mathbf{x}\in\R^{n}$ and output $\mathbf{x}'\in\R^{m}$ takes the form $\mathbf{x}' = \sigma(\mathbf{W}\mathbf{x}+\mathbf{b})$, where $\mathbf{W}\in\R^{m\times n}$ is the weight matrix, $\mathbf{b}\in\R^{m}$ the bias, and $\sigma$ the activation. The matrix product $\mathbf{W}\mathbf{x}$ can be evaluated on a quantum computer when $\mathbf{W}$ is orthogonal; the resulting layer is called a quantum layer. Because adding the bias and applying the nonlinearity remain classical, a quantum layer comprises three steps, illustrated in \cref{fig:quantum_layer}: (i) loading the classical input onto the quantum circuit, (ii) performing the orthogonal matrix multiplication with a pyramidal circuit, and (iii) reading the result back into a classical vector by tomography.

For the reconfigurable beam-splitter gate, the basic building block is the two-qubit reconfigurable beam-splitter (RBS) gate, whose action on the two-qubit subspace is
\begin{equation}
U_{\mathrm{RBS}}(\theta) =
\begin{pmatrix}
1 & 0 & 0 & 0\\
0 & \cos\theta & \sin\theta & 0\\
0 & -\sin\theta & \cos\theta & 0\\
0 & 0 & 0 & 1
\end{pmatrix}.
\label{eq:rbs}
\end{equation}
It rotates within the single-excitation subspace, $\lvert 01\rangle \mapsto \cos\theta\,\lvert 01\rangle - \sin\theta\,\lvert 10\rangle$ and $\lvert 10\rangle \mapsto \sin\theta\,\lvert 01\rangle + \cos\theta\,\lvert 10\rangle$, while leaving $\lvert 00\rangle$ and $\lvert 11\rangle$ fixed. Because every RBS gate preserves the number of excitations, an entire circuit built from these gates stays within the unary subspace, which is what makes the loading and tomography below exact.

\subsubsection{Loading classical data input}
\label{sec:loading}
To operate on a classical vector $\mathbf{x}\in\R^{n}$, it must first be written into a quantum state, a step called data loading. The probabilistic nature of quantum mechanics requires $\lVert\mathbf{x}\rVert_2=1$. To guarantee this without discarding the scale of $\mathbf{x}$, an extra dimension is appended at the first layer: each element is rescaled to $[-1,1]$ and the new coordinate is set to $\sqrt{1-\sum_i x_i^2/n}$, transforming $\mathbf{x}$ into the unit-norm vector
\begin{equation}
\Big(x_1,\ x_2,\ \dots,\ x_n,\ \sqrt{\textstyle 1-\sum_i x_i^2/n}\,\Big)^{\!\top}.
\label{eq:normload}
\end{equation}
For subsequent layers the vector is already unit-norm and is simply divided by $\lVert\mathbf{x}\rVert_2$. The circuit is initialized with the first qubit in $\lvert 1\rangle$ and the rest in $\lvert 0\rangle$, and a cascade of RBS gates with angles $\alpha_1,\dots,\alpha_{n-1}$, where $\alpha_1=\arccos(x_1)$, $\alpha_2=\arccos\!\big(x_2\sin^{-1}\alpha_1\big)$, and so on, produces the unary state
\begin{equation}
\lvert \mathbf{x}\rangle = x_1\lvert 10\cdots0\rangle + x_2\lvert 01\cdots0\rangle + \cdots + x_n\lvert 00\cdots1\rangle,
\label{eq:unary}
\end{equation}
in which the amplitude of the $j$th unary basis state $\lvert e_j\rangle$ is exactly $x_j$. When the input and output dimensions differ, the circuit uses $\max(m,n)$ qubits: for $m<n$ the data occupies all $n$ qubits, while for $m>n$ it is loaded onto the bottom $n$ qubits and the upper $m-n$ are left in $\lvert 0\rangle$.

\subsubsection{Quantum pyramidal circuit}
\label{sec:qorthonn}
Once $\mathbf{x}$ is loaded as a unary state, the product $\mathbf{y}=\mathbf{W}\mathbf{x}$
is performed in place by a pyramidal
circuit. The construction rests
on the fact that any $\mathbf{W}\in SO(n)$ factors into $n(n-1)/2$ planar
(Givens) rotations, each acting on a single coordinate pair. Comparing this with
\cref{eq:rbs}, an RBS gate acting on qubits $j$ and $j+1$ implements exactly
such a rotation on the amplitudes of $\lvert e_j\rangle$ and $\lvert
e_{j+1}\rangle$ and leaves every other unary amplitude untouched. A product of
RBS gates therefore realizes a product of Givens rotations, and the orthogonal
matrix is represented not by its $n^2$ entries but by the angles of the gates
that generate it.
 
The gates are arranged in a pyramid: for the square case $m=n$ the first
diagonal of the pyramid couples qubit pairs $(1,2),(3,4),\dots$, the second
couples $(2,3),(4,5),\dots$, and the pattern alternates until every pair has
been coupled through some path, using $d=n(n-1)/2$ angles
$\theta_1,\dots,\theta_d$ in total. Writing $C_{\theta}=\cos\theta$ and
$S_{\theta}=\sin\theta$, the $n=4$ pyramid realizes
\begin{equation}
\mathbf{W} =
\underbrace{\begin{pmatrix} C_{\theta_1} & S_{\theta_1} & & \\ -S_{\theta_1} & C_{\theta_1} & & \\ & & 1 & \\ & & & 1\end{pmatrix}}_{\text{RBS}(\theta_1)\ \text{on}\ (1,2)}
\cdots
\underbrace{\begin{pmatrix} 1 & & & \\ & C_{\theta_5} & S_{\theta_5} & \\ & -S_{\theta_5} & C_{\theta_5} & \\ & & & 1\end{pmatrix}}_{\text{RBS}(\theta_5)\ \text{on}\ (2,3)}
\underbrace{\begin{pmatrix} C_{\theta_6} & S_{\theta_6} & & \\ -S_{\theta_6} & C_{\theta_6} & & \\ & & 1 & \\ & & & 1\end{pmatrix}}_{\text{RBS}(\theta_6)\ \text{on}\ (1,2)},
\label{eq:pyramid_factors}
\end{equation}
the omitted factors following the same alternating pattern. Applied to the
loaded state, the circuit produces
\begin{equation}
\lvert \mathbf{y}\rangle = \lvert \mathbf{W}\mathbf{x}\rangle
= \sum_{i,j} W_{ji}\,x_i \,\lvert e_j\rangle,
\label{eq:pyramid_action}
\end{equation}
so the $j$th unary amplitude of the output state is exactly the $j$th component
of $\mathbf{W}\mathbf{x}$. Because every RBS gate conserves excitation number,
the state never leaves the unary subspace, and the entire $2^n$-dimensional
Hilbert space is used only as a carrier for an $n$-dimensional vector. This
sparsity is what makes the tomography of \cref{sec:tomography} affordable.
 
When $m\neq n$ the pyramid is truncated to the shape appropriate to the
rectangular map, and the number of free angles is given by \cref{eq:nangles}.
Two consequences of that formula matter for the present work. First, the angle
count is governed by $\min(m,n)$: a layer whose input is low-dimensional has few
parameters no matter how wide its output, because a $\min(m,n)$-dimensional
subspace requires only that many rotations to place inside the larger space.
Second, for $|m-n|\leq 1$ the pyramid geometry coincides with the square case,
though the layer as a whole still differs through the loading and tomography stages. When we replace a raw coordinate by a spectral embedding in
\cref{sec:qsedonet}, it is precisely $\min(m,n)$ that increases, from the
coordinate dimension to the embedding dimension, while
$\max(m,n)$, and therefore the qubit count of \cref{eq:qubits}, is unchanged.
 
Stacking such layers, each followed by a classical bias and nonlinearity, yields
a quantum orthogonal neural network (QOrthoNN). Because the pyramid shares its
mathematical expression with a classical orthogonal layer, a QOrthoNN inherits
the stable-gradient and conditioning benefits of
orthogonality while admitting the
quantum evaluation above. The full single-layer circuit of \cref{fig:quantum_layer}
contains at most $3n+O(1)$ RBS gates, at most $2n+1$ for the combined loader,
pyramid, and inverse loader, plus $n-1$ for the final loader and a constant
number of ancilla operations, so after transpilation to a hardware basis, the
depth remains $O(n)$ in the layer width.

\subsubsection{Tomography for extracting classical output}
\label{sec:tomography}
After the pyramid, the output vector $\mathbf{y}=\mathbf{W}\mathbf{x}$ resides
in the amplitudes of a unary state and must be returned to classical form so
that the bias and nonlinearity can be applied. Full quantum state tomography is
prohibitively expensive in
general, but the unary encoding
makes the problem tractable: only $r=\max(m,n)$ amplitudes are nonzero, and each
is the amplitude of a single computational basis state, so the magnitudes
$\lvert y_j\rvert$ follow directly from the measurement probabilities of those
$r$ basis states.
 
Magnitudes alone are insufficient, since a neural network layer requires signed
outputs. The procedure adopted here recovers the signs by interference against a known reference. As shown in
\cref{fig:quantum_layer}, an ancillary qubit is placed in superposition by a
Hadamard gate and entangled with the first register qubit by a CNOT before the
loader. After the pyramid, the circuit applies the adjoint of the loader for the
uniform unit vector $\mathbf{1}/\sqrt{r}$, followed by a $X$ gate on the
ancilla and a second CNOT, then reloads $\mathbf{1}/\sqrt{r}$ and applies a
final Hadamard. The resulting state is
\begin{equation}
\frac{1}{2}\sum_j\Big(y_j+\tfrac{1}{\sqrt{r}}\Big)\lvert 0,e_j\rangle
\;+\;
\frac{1}{2}\sum_j\Big(y_j-\tfrac{1}{\sqrt{r}}\Big)\lvert 1,e_j\rangle ,
\label{eq:tomo_state}
\end{equation}
where $\lvert \xi,e_j\rangle$ denotes a state in which the ancilla is in state $\lvert\xi\rangle$ and the
register is in the $j$th unary state. The reference amplitude $1/\sqrt{r}$ is thus
added on one ancilla branch and subtracted on the other, and the sign of $y_j$
is exposed by which branch is more likely:
\begin{equation}
\Pr[0,e_j]-\Pr[1,e_j] = \frac{y_j}{\sqrt{r}}
\quad\Longrightarrow\quad
\operatorname{sign}(y_j)=
\begin{cases}
+1, & \Pr[0,e_j]\ge \Pr[1,e_j],\\[2pt]
-1, & \text{otherwise.}
\end{cases}
\label{eq:sign}
\end{equation}
The magnitude then follows from the corresponding branch probability.
\begin{equation}
y_j =
\begin{cases}
\operatorname{sign}(y_j)\Big(2\sqrt{\Pr[0,e_j]}-\tfrac{1}{\sqrt{r}}\Big), & y_j> 0,\\[4pt]
\operatorname{sign}(y_j)\Big(2\sqrt{\Pr[1,e_j]}+\tfrac{1}{\sqrt{r}}\Big), & y_j\le 0.
\end{cases}
\label{eq:value}
\end{equation}
Two properties of this scheme are worth stating. The estimation error of each
$y_j$ is independent of the register size $r$, because the reference amplitude
enters every component identically; and since the tomography circuit has depth
$O(n)$ and $O(1/\delta^2)$ measurements suffice for accuracy $\delta$, the whole
extraction costs $O(n/\delta^{2})$. When the output dimension is smaller than
the input dimension, $m<n$, the circuit is unchanged and only the bottom $m$
unary states are read, so $\lvert e_j\rangle$ in
\cref{eq:tomo_state,eq:sign,eq:value} refers to the $j$th unary state of the
bottom $m$ qubits. In the ideal infinite-shot limit the procedure returns
$\mathbf{y}$ exactly, so the quantum layer computes precisely the classical
orthogonal map; the residual differences observed in simulation, reported in
\cref{sec:results}, originate in the finite precision of the classical
arithmetic and of the loader angle reconstruction rather than in any modeling
difference.
 
For numerically stable loader angles, 
the reference implementation computes the loader angles of
\cref{sec:loading} by the recursion
$\alpha_i=\arccos\!\big(x_i\prod_{j<i}\sin^{-1}\alpha_j\big)$, with the final
angle obtained from a ratio of the last two components. Both operations divide
by quantities that vanish when components of $\mathbf{x}$ are zero, and the
released code compensates by perturbing near-zero entries by $10^{-7}$. This is
adequate for the smooth, dense activation vectors of the original examples, but
it fails for ReLU networks of moderate width, where a substantial fraction of
each activation vector is exactly zero: in our width-$10$ experiments between
four and seven of ten components vanish for every input, and the accumulated
product diverges. We therefore compute the angles from cumulative tail norms,
\begin{equation}
\alpha_i = \operatorname{atan2}\!\Big(\big\lVert (x_{i+1},\dots,x_n)\big\rVert_2,\ x_i\Big),
\qquad
\alpha_{n-1} = \operatorname{atan2}(x_n, x_{n-1}),
\label{eq:stable_loader}
\end{equation}
which involves no division, is exact for vanishing components, and returns the
signed angle directly. The prepared state is mathematically identical to
\cref{eq:unary}; only the numerics differ. On sparse test vectors the
reformulation reduces the round-trip error of an identity pyramid from failure
to $\sim\!10^{-14}$, and it is what makes the ReLU networks of
\cref{sec:results} simulable at all.


\subsubsection{Quantum orthogonal neural network}
\label{sec:qorthonn}
Once $\mathbf{x}$ is loaded, the orthogonal product $\mathbf{y}=\mathbf{W}\mathbf{x}$ is performed by a pyramidal circuit. An orthogonal matrix $\mathbf{W}$ is decomposed into a sequence of planar (Givens) rotations, each realized by one RBS gate, and the gates are arranged in a pyramid. For the square case $m=n$ the pyramid uses $d=n(n-1)/2$ angles $\theta_1,\dots,\theta_d$; for rectangular shapes the geometry adapts as shown in the top row of \cref{fig:quantum_layer}, giving in general
\begin{equation}
N_\theta = \tfrac{1}{2}\,(2\max - 1 - \min)\,\min, \qquad \max=\max(m,n),\ \min=\min(m,n).
\label{eq:nangles}
\end{equation}
Stacking such layers, each followed by a classical bias and nonlinearity, yields a quantum orthogonal neural network (QOrthoNN). Because it shares the same mathematical expression as a classical orthogonal network, a QOrthoNN inherits the stable-gradient and conditioning benefits of orthogonality while admitting the quantum evaluation above.

A key consequence for the present work is the qubit count. A layer with input dimension $n_{\mathrm{in}}$ and output dimension $n_{\mathrm{out}}$ requires
\begin{equation}
N_{\mathrm{qubits}} = 1 + \max(n_{\mathrm{in}}, n_{\mathrm{out}}),
\label{eq:qubits}
\end{equation}
one ancilla for the tomography plus $\max(n_{\mathrm{in}},n_{\mathrm{out}})$ qubits for the unary register. Since every hidden layer has width equal to the network width $W$, and $W$ is at least as large as any trunk input we consider, the trunk qubit count is fixed at $1+W$ regardless of the input embedding dimension, provided that dimension does not exceed $W$. This is the property Quantum SEDONet exploits.

\subsection{Quantum DeepONet}
\label{sec:qdeeponet}
DeepONet approximates the solution operator $\G$ of a PDE, mapping an input function $v$ (an initial condition or source term) to the solution $u$, through two subnetworks. The branch network encodes $v$ samples at $q$ sensor locations $\{z_1,\dots,z_q\}$ into a coefficient vector, and the trunk network encodes the query coordinate $\xi$ into a basis vector; their inner product gives
\begin{equation}
\G'_\theta(v)(\xi) = \sum_{k=1}^{p} b_k(v)\,t_k(\xi) + b_0,
\label{eq:deeponet}
\end{equation}
where $b_0$ is a scalar bias and $\theta$ collects the trainable parameters. Quantum DeepONet replaces the branch and trunk multilayer perceptrons with QOrthoNNs: each subnetwork becomes a stack of quantum layers of the form above, with biases and nonlinearities applied classically between them. In the ideal regime this leaves the learned operator unchanged relative to a classical orthogonal DeepONet, while offering an asymptotic inference speed-up per layer.

\subsection{Quantum SEDONet}
\label{sec:qsedonet}
Quantum SEDONet retains the Quantum DeepONet architecture in full and modifies
only how the query coordinate enters the trunk. The motivation is a
representational bottleneck. In the released architectures the trunk receives its
coordinate input completely raw, so each coordinate reaches the trunk as a single
linear input and the trunk QOrthoNN must synthesize every oscillation of the
solution in that coordinate through its nonlinearities. Coordinate networks are
biased toward low
frequencies, so
the approximation error concentrates where the solution varies fastest;
empirically, for the Burgers data the mean absolute error at $t=0$ exceeds that
at $t>0.2$ by a factor of roughly twenty, and the single $t=0$ slice accounts for
about ten times the squared error of all interior slices combined.

\begin{figure}[H]
\centering
\includegraphics[width=\textwidth]{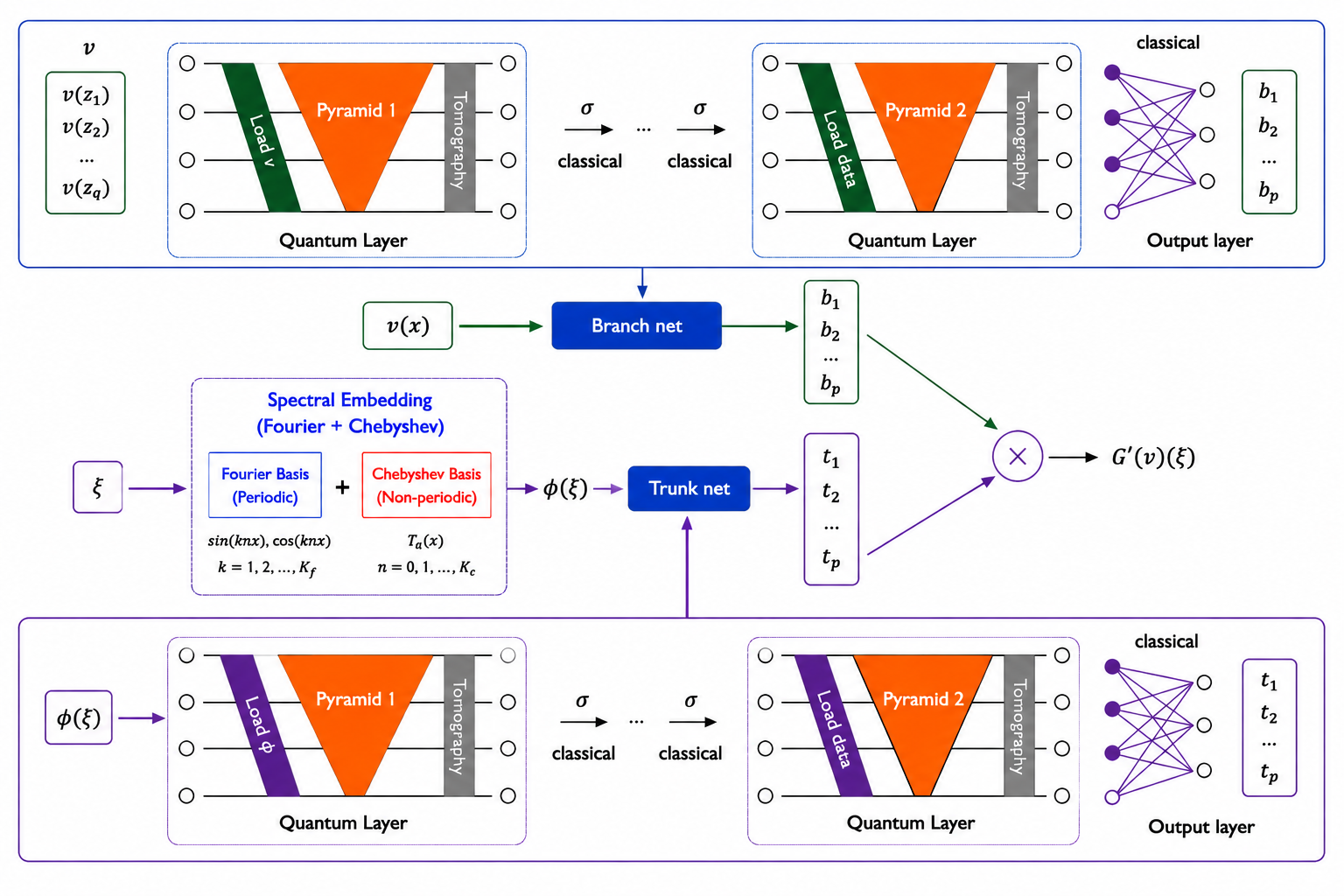}
\caption{The Quantum SEDONet architecture. As in Quantum DeepONet, a branch net encodes the input function $v$ and a trunk net encodes the query coordinate $\xi$; each subnetwork is a stack of QOrthoNN quantum layers (Load $\to$ Pyramid $\to$ Tomography, with classical $\sigma$ between layers and a classical output layer), and the two outputs are combined by an inner product to give $\G'(v)(\xi)$. Quantum SEDONet adds a single new block, the highlighted spectral embedding, which maps the trunk coordinate $\xi$ to $\phi(\xi)$ using a Fourier basis for periodic coordinates and a Chebyshev basis for bounded non-periodic ones before the quantum layers.}
\label{fig:flowchart}
\end{figure}

Quantum SEDONet inserts a spectral embedding on the trunk input, replacing the query coordinate $\xi$ with a feature vector $\phi(\xi)$ before the first trunk layer. Each coordinate is expanded in the classical spectral basis matched to its boundary condition. A periodic coordinate $x$ receives the Fourier features
\begin{equation}
\big[\cos(2\pi k x),\ \sin(2\pi k x)\big]_{k=1}^{K_f},
\label{eq:fourier}
\end{equation}
which respect the periodicity exactly, and a bounded, non-periodic coordinate $t\in[0,1]$ receives the Chebyshev features
\begin{equation}
\big[T_1(2t-1),\ T_2(2t-1),\ \dots,\ T_{K_c}(2t-1)\big],
\label{eq:cheb}
\end{equation}
where $T_n$ denotes the Chebyshev polynomials of the first kind. Two details, the first being that the constant polynomial $T_0$ is omitted: it would produce a degenerate column that the min-max normalization of \cref{eq:normload} cannot rescale. Second, $T_1(2t-1)=2t-1$ is affine in $t$, so the Chebyshev embedding strictly contains the original raw coordinate as a special case; the network can recover the baseline behavior if that were optimal, and the embedding can therefore never be representationally worse. Chebyshev polynomials provide a natural global basis on bounded, non-periodic intervals and can efficiently represent endpoint-sensitive structure. For a problem whose only coordinate is bounded and nonperiodic, a steady one-dimensional problem, for instance, the embedding is Chebyshev alone with no Fourier term, since imposing a Fourier basis would falsely assume periodicity of a solution satisfying Dirichlet conditions. The governing principle is simply that each coordinate is expanded in the classical spectral basis appropriate to its boundary condition. \Cref{fig:flowchart} shows the full architecture with this embedding highlighted.

The essential property is that this embedding is free in the resource that constrains the platform. The feature vector $\phi(\xi)$ has dimension at most $W-1$, where $W$ is the network width, so by \cref{eq:qubits} the first trunk layer requires $1+\max(\dim\phi, W)=1+W$ qubits, identical to the raw-input baseline. Every subsequent hidden layer is width-to-width and unchanged. Consequently the qubit count, the circuit depth, and the number of evaluated circuits are all identical between Quantum DeepONet and Quantum SEDONet; the only overhead is the additional RBS angles in the first trunk layer, which raise the total parameter count by a few percent (\cref{tab:cost}). This is the sense in which Quantum SEDONet is not a trade of accuracy against quantum resources but a strict improvement in the regime relevant to near-term hardware.


Algorithm~\ref{alg:qsedonet_infer} gives the full inference procedure. The
spectral embedding is applied as classical preprocessing before the first quantum
layer, so the register allocated at each layer is governed by
Eq.~\eqref{eq:qubits} exactly as it would be for the raw-coordinate baseline, and
the qubit count appears explicitly at the point of allocation rather than being
absorbed into the architecture description. Every subsequent step, the unary
loading, the pyramid, and the tomography, is identical in both arms, which is
what makes the two procedures directly comparable circuit for circuit.
\begin{algorithm}[ht!]
\caption{Quantum inference for Quantum SEDONet}
\label{alg:qsedonet_infer}
\begin{algorithmic}[1]
\Require Trained angles $\theta$, biases $b$, output bias $b_0$; query coordinate
         $\xi$; input function $v$ at $q$ sensors; layer widths
         $\{(n_{\mathrm{in}}^{(\ell)},n_{\mathrm{out}}^{(\ell)})\}_{\ell=1}^{L}$
\Ensure Prediction $\mathcal{G}'_\theta(v)(\xi)$

\Function{QuantumLayer}{$\mathbf{x},\theta^{(\ell)},n_{\mathrm{in}},n_{\mathrm{out}}$}
    \State $r\gets\max(n_{\mathrm{in}},n_{\mathrm{out}})$;
           allocate $1+r$ qubits (one ancilla $+$ unary register)
           \Comment{Eq.~\eqref{eq:qubits}}
    \If{$\ell=1$} \Comment{first layer only}
        \State rescale each entry of $\mathbf{x}$ to $[-1,1]$ and append
               $\sqrt{1-\sum_i x_i^2/n}$ \Comment{Eq.~\eqref{eq:normload}}
    \Else
        \State $\mathbf{x}\gets\mathbf{x}/\lVert\mathbf{x}\rVert_2$
    \EndIf
    \State \textbf{Load:} $\alpha_i\gets\mathrm{atan2}\big(
           \lVert(x_{i+1},\dots,x_n)\rVert_2,\;x_i\big)$,
           $\alpha_{n-1}\gets\mathrm{atan2}(x_n,x_{n-1})$
           \Comment{Eq.~\eqref{eq:stable_loader}; division-free, exact at $x_i=0$}
    \State prepare $\lvert x\rangle=\sum_j x_j\lvert e_j\rangle$ by the RBS cascade
           $\{\alpha_i\}$ \Comment{Eq.~\eqref{eq:unary}}
    \State \textbf{Pyramid:} apply the $N_\theta=\tfrac12(2r-1-\min)\min$ RBS gates
           of $\theta^{(\ell)}$, giving $\lvert y\rangle=\lvert\mathbf{W}\mathbf{x}\rangle$
           \Comment{Eqs.~\eqref{eq:nangles},~\eqref{eq:pyramid_action}}
    \State \textbf{Tomography:} $H$ and CNOT on the ancilla before the loader;
           after the pyramid apply the adjoint loader of $1/\sqrt{r}$, $X$ on the
           ancilla, a second CNOT, reload $1/\sqrt{r}$, and a final $H$
           \Comment{Eq.~\eqref{eq:tomo_state}}
    \State recover $\mathrm{sign}(y_j)$ from
           $\Pr[0,e_j]-\Pr[1,e_j]$ and $\lvert y_j\rvert$ from the corresponding
           branch probability \Comment{Eqs.~\eqref{eq:sign},~\eqref{eq:value}}
    \State \Return $\sigma\big(\mathbf{y}[1{:}n_{\mathrm{out}}]+b^{(\ell)}\big)$
           \Comment{bias and nonlinearity applied classically}
\EndFunction

\Statex
\State \textbf{Embed:} $\phi(\xi)\gets$ boundary-matched Fourier/Chebyshev features
       (Stage 1 of Alg.~\ref{alg:qsedonet_train})
       \Comment{classical preprocessing; no qubit cost}
\State $\mathbf{h}\gets\phi(\xi)$
\For{$\ell=1,\dots,L$} \Comment{trunk}
    \State $\mathbf{h}\gets\Call{QuantumLayer}{\mathbf{h},
           \theta_{\mathrm{trunk}}^{(\ell)},n_{\mathrm{in}}^{(\ell)},n_{\mathrm{out}}^{(\ell)}}$
\EndFor
\State $\mathbf{t}(\xi)\gets$ classical output layer applied to $\mathbf{h}$
\State $\mathbf{b}(v)\gets$ same loop with $\theta_{\mathrm{branch}}$ on the sensor
       vector $v$
\State \Return $\mathcal{G}'_\theta(v)(\xi)=\mathbf{b}(v)^{\top}\mathbf{t}(\xi)+b_0$
\end{algorithmic}
\end{algorithm}

\subsection{Training quantum SEDONet}
\label{sec:training}
Following this, the network is trained classically. The pyramidal circuit shares the same mathematical expression as a classical orthogonal neural network, so the angles $\theta$ are optimized on a classical computer by backpropagation, and the trained angles are then transferred to the quantum circuit for evaluation. We minimize the mean squared error between the predicted and reference solution values over a set of input functions and query points arranged as a Cartesian product,
\begin{equation}
\mathcal{L}(\theta) = \frac{1}{N}\sum_{i=1}^{N}\big\lVert \G'_\theta(v_i) - u_i \big\rVert_2^2,
\label{eq:loss}
\end{equation}
using the Adam optimizer~\cite{kingma2015adam}. Only the trunk input map differs between the Quantum DeepONet and Quantum SEDONet arms; all other settings, namely width, depth, learning rate, iteration count, and training data, are held fixed, so any difference in accuracy is attributable to the spectral embedding alone. After training, the learned angles are loaded into the tomography pipeline of \cref{sec:tomography} and we verify on a subset of test functions that the ideal-quantum prediction matches the classical prediction to within $10^{-8}$, confirming correct parameter transfer.

Algorithm~\ref{alg:qsedonet_train} collects the complete procedure. The
boundary-matched embedding is precomputed once for the fixed trunk grid
(Stage~1), and the check $\dim\phi \le W-1$ is what makes the zero-qubit-overhead
claim a property of the procedure rather than of a particular benchmark.
\begin{algorithm}[ht!]
\caption{Training procedure for Quantum SEDONet}
\label{alg:qsedonet_train}
\begin{algorithmic}[1]
\Require Training set $\mathcal{D}=\{(v^{(i)},u^{(i)})\}_{i=1}^{N}$ with $v^{(i)}$
         sampled at $q$ sensors; coordinate grid $\{\xi^{r}\}_{r=1}^{R}$,
         $\xi=(\xi_1,\dots,\xi_d)$; boundary-condition tags
         $\{\mathrm{bc}(\xi_j)\}_{j=1}^{d}$; network width $W$; number of trunk
         channels $p$; learning rate $\eta$; iteration count $M$
\Ensure Trained RBS angles $\theta=(\theta_{\mathrm{branch}},\theta_{\mathrm{trunk}})$,
        biases $b$, output bias $b_0$

\Statex \hspace{-1.2em}\textbf{Stage 1: boundary-matched spectral embedding (precomputed once)}
\State Allocate the feature budget so that $\dim\phi \le W-1$, giving per-coordinate
       orders $K_f$ (Fourier) and $K_c$ (Chebyshev)
\For{each trunk coordinate $\xi_j$, $j=1,\dots,d$}
    \If{$\mathrm{bc}(\xi_j)$ is periodic}
        \State $\phi_j(\xi_j)\gets\big\{\cos(2\pi k\xi_j),\,\sin(2\pi k\xi_j)\big\}_{k=1}^{K_f}$
               \Comment{Eq.~\eqref{eq:fourier}}
    \Else \Comment{bounded, non-periodic (Dirichlet or initial condition)}
        \State $\phi_j(\xi_j)\gets\big\{T_n(2\xi_j-1)\big\}_{n=1}^{K_c}$, omitting $T_0$
               \Comment{Eq.~\eqref{eq:cheb}}
    \EndIf
\EndFor
\State $\phi(\xi)\gets\big[\phi_1(\xi_1),\dots,\phi_d(\xi_d)\big]$;
       evaluate $\phi(\xi^{r})$ for all $r=1,\dots,R$ and cache
\State \textbf{assert} $\dim\phi \le W-1$
       \Comment{guarantees $1+\max(\dim\phi,W)=1+W$ qubits, Eq.~\eqref{eq:qubits}}

\Statex \hspace{-1.2em}\textbf{Stage 2: classical optimization of the orthogonal parameterization}
\State Initialize $\theta_{\mathrm{branch}},\theta_{\mathrm{trunk}},b,b_0$;
       set branch layer widths $[q+1,W,\dots,W]$ and trunk widths $[\dim\phi,W,\dots,W]$
\For{$m=1,\dots,M$}
    \For{each mini-batch $\mathcal{B}\subset\mathcal{D}$}
        \State \textbf{Branch:} $\mathbf{b}(v^{(i)})\gets
               \mathrm{OrthoNN}(\theta_{\mathrm{branch}};v^{(i)})\in\mathbb{R}^{p}$
               for $i\in\mathcal{B}$
        \State \textbf{Trunk:} $\mathbf{t}(\xi^{r})\gets
               \mathrm{OrthoNN}(\theta_{\mathrm{trunk}};\phi(\xi^{r}))\in\mathbb{R}^{p}$
               \Comment{cached $\phi$, not raw $\xi$}
        \State \textbf{Synthesis:} $\widehat{u}^{(i)}(\xi^{r})\gets
               \mathbf{b}(v^{(i)})^{\top}\mathbf{t}(\xi^{r})+b_0$
               \Comment{Eq.~\eqref{eq:deeponet}}
        \State \textbf{Loss:} evaluate $\mathcal{L}(\theta)$
               \Comment{Eq.~\eqref{eq:loss}}
        \State \textbf{Update:} one Adam step with learning rate $\eta$ on
               $(\theta,b,b_0)$
    \EndFor
\EndFor
\State \Return the final iterate $(\theta,b,b_0)$
       \Comment{no checkpoint selection; both arms train for identical $M$}
\end{algorithmic}
\end{algorithm}

%
%
\section{Results}
\label{sec:results}

\subsection{Experimental setup}
\label{sec:setup}
We evaluate on four benchmarks: the antiderivative, advection, and Burgers
problems from the Quantum DeepONet study, and a channel
Poisson problem introduced here to exercise both spectral bases on a single
geometry. Together they span the full range of the basis-selection rule. The
antiderivative has a single bounded, non-periodic coordinate and therefore
receives a Chebyshev expansion alone; advection and Burgers pair a periodic
spatial coordinate with a nonperiodic temporal one; and channel Poisson draws
both bases from purely spatial boundary conditions. Input functions are drawn
from Gaussian random fields~\cite{rasmussen2006gaussian}, and reported errors are
the mean relative $L^2$ error over the full test set: $100$ functions for the
antiderivative on a $30$-point grid, $100$ for Burgers and $200$ for advection on
a $50\times50$ space--time grid, and $100$ for channel Poisson on a $40\times40$
spatial grid.

For each benchmark we train a baseline, the released trunk representation which
we call the Quantum DeepONet arm, and a Quantum SEDONet arm differing only in the
trunk input map; width, depth, learning rate, iteration count, and training data
are held fixed, so any difference in accuracy is attributable to the embedding
alone. Both arms use orthogonal branch and trunk networks trained with Adam for
the same fixed iteration count, and the final iterate is evaluated. After training
we transfer the learned angles into the quantum tomography pipeline of
\cref{sec:tomography} and confirm that the ideal-quantum prediction matches the
classical prediction to within the precision of the classical arithmetic.
Accuracy and resource comparisons are collected in
\cref{tab:main,tab:cost} of \cref{sec:summary}.

\subsection{Function approximation}
\label{sec:function_approx}
Before the operator benchmarks we reproduce the two function-approximation tests. These use a bare QOrthoNN with a single scalar
input and no branch-trunk structure, so no spectral trunk embedding applies and
no Quantum SEDONet arm exists; their purpose is to validate the
classical-to-quantum parameter transfer and the tomography pipeline end to end
before those components are relied upon in a two-network architecture.

The first target is the Runge function $f(x)=1/(1+25x^2)$ on $[-1,1]$,
approximated by an OrthoNN of layer sizes $[2,3,3,1]$ with $\tanh$ activations,
trained on $80$ points, and tested on $100$, using Adam at a learning rate
$10^{-4}$ for $5\times10^{4}$ iterations. The $\tanh$ activation is used in place
of ReLU because the width-$3$ network is acutely vulnerable to the dying ReLU
problem~\cite{lu2020dying}. The second target is $f(x)=\sum_{k=1}^{4}\sin(kx)$
on $[-\pi,\pi]$, approximated by an OrthoNN of layer sizes $[2,10,10,10,1]$ with
ReLU activations, trained on $200$ points and tested on $100$, at a learning rate
$5\times10^{-4}$ for $4\times10^{4}$ iterations. In both cases the scalar input
is augmented by the norm-carrying dimension of \cref{eq:normload} before
entering the first quantum layer.

\begin{figure}[H]
\centering
\begin{subfigure}[b]{0.48\textwidth}
    \centering
    \includegraphics[width=\textwidth]{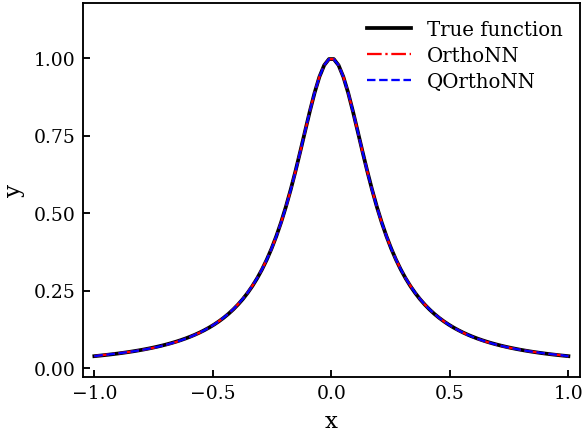}
    \caption{}
\end{subfigure}
\hfill
\begin{subfigure}[b]{0.48\textwidth}
    \centering
    \includegraphics[width=\textwidth]{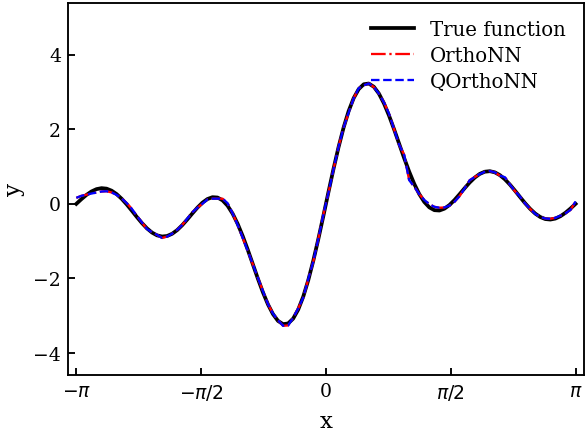}
    \caption{}
\end{subfigure}
\caption{Ideal quantum simulation of function approximation with a QOrthoNN.
Black, red, and blue denote the reference function, the classical OrthoNN
prediction, and the ideal quantum simulation of the QOrthoNN respectively; the
three curves are superimposed by construction.
(\textbf{a}) $f(x)=1/(1+25x^{2})$ on $[-1,1]$, relative $L^2$ error $0.136\%$,
classical--quantum agreement $7.8\times10^{-9}$.
(\textbf{b}) $f(x)=\sum_{k=1}^{4}\sin(kx)$ on $[-\pi,\pi]$, relative $L^2$ error
$3.308\%$, agreement $1.0\times10^{-6}$.}
\label{fig:function_approx}
\end{figure}

\Cref{fig:function_approx} shows the results. The classical OrthoNN and the
ideal quantum simulation of the corresponding QOrthoNN are visually
indistinguishable from each other and from the reference function, which is the
expected outcome: they are the same network evaluated in two different places.
Quantitatively we obtain a relative $L^2$ error of $0.136\%$ for the Runge
function and $3.308\%$ for the sinusoidal sum, with the classical and quantum
predictions agreeing to $7.8\times10^{-9}$ and $1.0\times10^{-6}$ respectively.
The agreement is limited by the precision of the classical arithmetic rather
than by the quantum simulation: the residual scales with the magnitude of the
network output, which is of order unity in the first case and of order four in
the second.

\subsection{Antiderivative operator}
\label{sec:antiderivative}
The advection, Burgers, and channel Poisson benchmarks all combine a periodic
coordinate with a non-periodic one, so each exercises both halves of the basis
rule simultaneously. The antiderivative operator isolates the Chebyshev half. We
learn
\begin{equation}
\frac{du}{dx}=v(x),\qquad x\in[0,1],\qquad u(0)=0,
\label{eq:antiderivative}
\end{equation}
that is, the operator $\G:v\mapsto u$ mapping a source to its antiderivative.
The single trunk coordinate is bounded and non-periodic, with an initial
condition at $x=0$ and no periodicity anywhere, so the rule of
\cref{sec:qsedonet} prescribes a Chebyshev expansion with no Fourier
term. This is the case anticipated in \cref{sec:qsedonet} but not previously
tested: a problem whose only coordinate is Dirichlet-type, where imposing a
Fourier basis would assert a periodicity the solution does not possess.

Input functions are drawn from a Gaussian random field with an RBF kernel of
length scale $\ell=1.0$, sampled on a dense grid of
$1000$ points, and reduced to $10$ branch sensors; reference solutions are
obtained by backward Euler integration of \cref{eq:antiderivative} on the dense
grid and evaluated at $30$ trunk points. We use $200$ training and $100$ test
functions. Both arms use a branch network of layer sizes $[11,10,10]$, ten
sensors plus the norm-carrying dimension, with ReLU activations, trained with
Adam at a learning rate $10^{-3}$ for $3\times10^{4}$ iterations, matching the
configuration of Quantum DeepONet for this problem. The only difference is in the trunk: the baseline receives the raw coordinate, giving layer sizes
$[2,10,10]$, while Quantum SEDONet receives it $T_1,\dots,T_9$ evaluated at $2x-1$,
giving $[10,10,10]$.

The qubit accounting is the same as elsewhere. By \cref{eq:qubits} the first
trunk layer needs $1+\max(2,10)=11$ qubits in the baseline and
$1+\max(10,10)=11$ with the embedding, so the trunk register is unchanged and
every subsequent layer is width-to-width. By \cref{eq:nangles} the trunk pyramid
grows from $17$ to $45$ angles, since the input subspace it must rotate is now
genuinely ten-dimensional rather than two-dimensional.

\begin{figure}[H]
\centering
\begin{subfigure}[b]{0.48\textwidth}
    \centering
    \includegraphics[width=\textwidth]{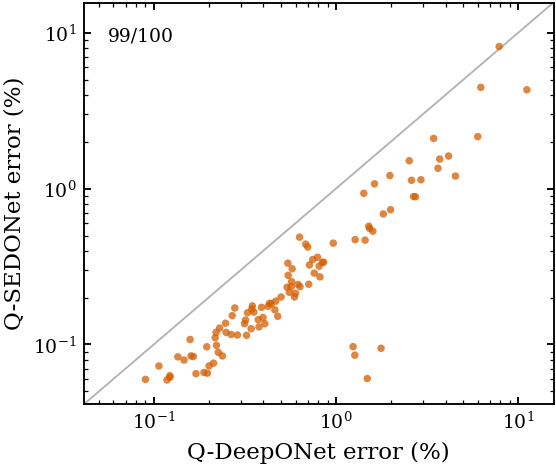}
    \caption{}
\end{subfigure}
\hfill
\begin{subfigure}[b]{0.48\textwidth}
    \centering
    \includegraphics[width=\textwidth]{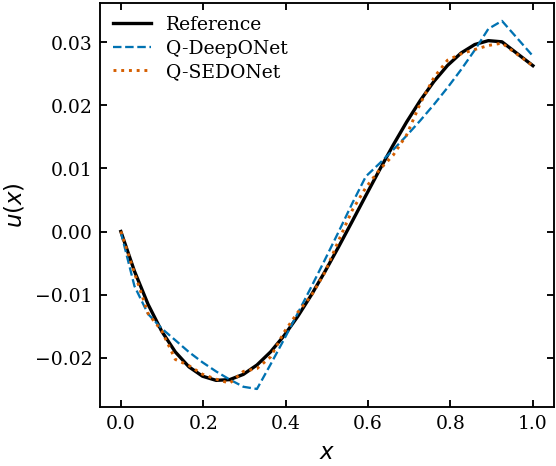}
    \caption{}
\end{subfigure}
\caption{Antiderivative operator, both arms at $11$ trunk qubits.
(\textbf{a}) Per-function comparison over the $100$ test functions, each point
giving the relative $L^2$ error of the two arms on the same input function;
points below the diagonal favor Quantum SEDONet, which is more accurate on $99$
of $100$ functions.
(\textbf{b}) The test function on which the baseline performs worst. The Quantum
DeepONet prediction is piecewise linear with visible hinge points, the signature
of a ReLU trunk driven by a single affine coordinate; the Chebyshev embedding
removes them. This sample has roughly $20\times$ smaller amplitude than a
typical test function, which inflates its relative error.}
\label{fig:antiderivative}
\end{figure}

The mean relative $L^2$ error falls from $1.159\%$ to $0.532\%$, a reduction of
$54.1\%$, the largest of the four benchmarks. The median error falls further in
relative terms, from $0.546\%$ to $0.187\%$, so the improvement is concentrated
on typical rather than exceptional inputs. \Cref{fig:antiderivative}(a) makes
the paired comparison explicit: each point is one test function, and $99$ of
$100$ lie below the diagonal. The points span nearly three decades of difficulty, and the improvement
persists across that entire range rather than being driven by any subset.

\Cref{fig:antiderivative}(b) shows the mechanism on the baseline's worst case.
The predicted solution is visibly piecewise linear, with kinks near
$x\approx0.33$, $0.55$, $0.65$, and $0.92$: a ReLU trunk whose input is a single
affine coordinate can only assemble a smooth oscillation from a small number of
hinges, and with a width of ten, it has few to spend. The Chebyshev features track
the reference almost exactly over the same interval. The sample shown has an
amplitude roughly twenty times smaller than a typical test function, which
inflates its relative error and is why it ranks worst for the baseline;
the absolute discrepancy is small in both arms, but the qualitative failure mode
is clearest here.

\subsection{Advection Equation}
\label{sec:advection}
Consider the one-dimensional advection equation
\begin{equation}
\frac{\partial u}{\partial t} + \frac{\partial u}{\partial x} = 0,
\qquad x \in [0,1], \quad t \in [0,1],
\label{eq:advection}
\end{equation}
with initial condition $u(x,0) = u_0(x)$ and periodic boundary conditions in
$x$. The operator to be learned maps the initial condition to the solution at
all later times,
\begin{equation}
\mathcal{G} : u_0(x) \mapsto u(x,t).
\label{eq:advection_operator}
\end{equation}
The initial conditions $u_0(x)$ are sampled from a Gaussian random field with an
exponential sine-squared kernel,
\begin{equation}
k(x_i, x_j) = \exp\!\left(-\frac{2\sin^2\!\big(\pi d(x_i,x_j)/p\big)}{\ell^2}\right),
\label{eq:expsinesq}
\end{equation}
where $d(x_i,x_j)$ is the distance between sample points, $p$ is the kernel
periodicity, and $\ell$ is the length scale. Following the reference setup for
this benchmark we take $p=1$, which matches the kernel period to the spatial
domain so that every sampled initial condition satisfies the periodic boundary
condition exactly, and $\ell=1.5$. Reference solutions are obtained from the
released data generator for this benchmark, which integrates
\cref{eq:advection} numerically. The branch network receives $u_0$ at $20$
uniformly spaced sensors, and the trunk is evaluated on a $50 \times 50$ grid
covering $x$ and $t$. Both arms are trained on identical data, so the comparison
isolates the trunk embedding.

\begin{figure}[H]
\centering
\includegraphics[width=\textwidth]{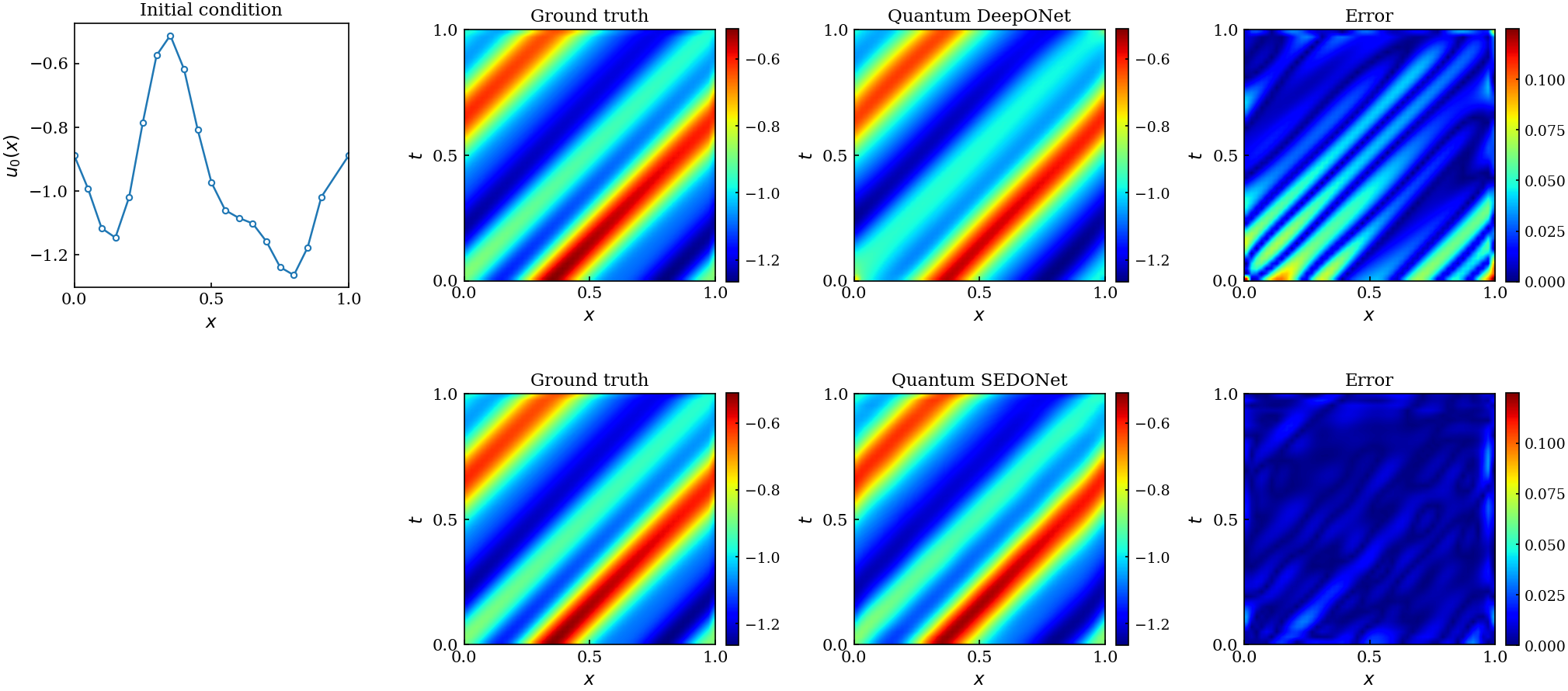}
\caption{Prediction comparison for one representative advection test sample,
showing the initial condition, the ground truth, Quantum DeepONet and Quantum
SEDONet predictions, and their corresponding absolute error maps on a common
color scale. The baseline error follows the diagonal transport characteristics;
the spectral embedding removes it.}
\label{fig:advection_fields}
\end{figure}

\Cref{fig:advection_fields} shows a representative advection test function. The
ground-truth solution is a wave transported along the diagonal characteristics
$x-t=\text{const}$. The baseline (top row) reproduces the wave but leaves a
structured error concentrated along those same diagonals: the raw-coordinate
trunk cannot represent the moving front sharply, so it smears energy along the
direction of transport. Quantum SEDONet (bottom row) removes almost all of this
structure, leaving a nearly featureless error field of markedly smaller
magnitude. The mechanism is transparent: for a pure translation the solution
operator acts on each spatial Fourier mode as a phase rotation, so a Fourier
embedding of the periodic coordinate $x$ represents the operator's action
essentially exactly, and the Chebyshev embedding of $t$ supplies the temporal
resolution the raw coordinate lacked.

\begin{figure}[H]
\centering
\includegraphics[width=0.5\textwidth]{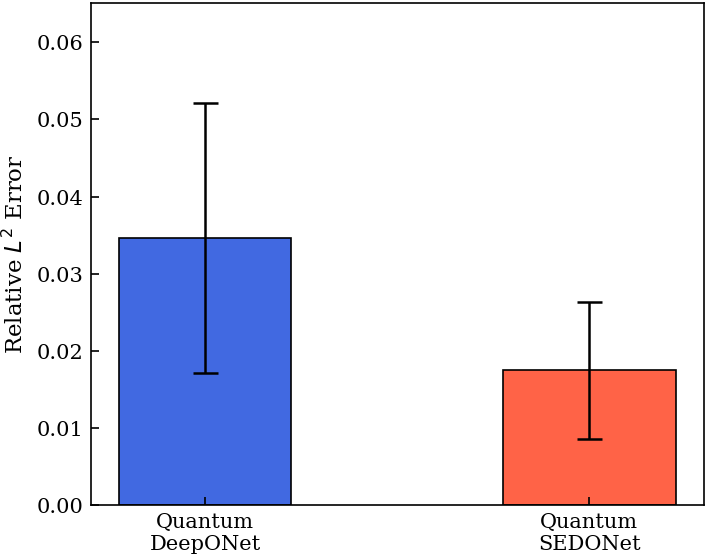}
\caption{Advection: mean relative $L^2$ error over $200$ test functions, with
standard-deviation bars. Quantum SEDONet improves on the baseline for $199$ of
$200$ functions (paired $t=23.3$).}
\label{fig:advection_bar}
\end{figure}

\Cref{fig:advection_bar} shows the aggregate comparison over all $200$ advection
test functions. The mean relative $L^2$ error falls from $3.463\%$ to $1.744\%$,
a $49.6\%$ reduction. The error bars show the standard deviation across test
functions and are necessarily wide, because problem difficulty varies from one
input to another; the meaningful comparison is paired, since both models are
evaluated on the same functions. Under that pairing, Quantum SEDONet is more
accurate on $199$ of $200$ functions, with a paired $t$-statistic of $23.3$.

\subsection{Burgers Equation}
\label{sec:burgers}
Building on the linear advection case, we next consider the nonlinear
one-dimensional Burgers equation
\begin{equation}
\frac{\partial u}{\partial t} + u\frac{\partial u}{\partial x}
= \nu \frac{\partial^2 u}{\partial x^2},
\qquad x \in [0,1], \quad t \in [0,1],
\label{eq:burgers}
\end{equation}
with initial condition $u_0(x)$, periodic boundary conditions in $x$, and
viscosity $\nu = 0.05$. The operator to be learned again maps the initial
condition to the solution at all later times,
\begin{equation}
\mathcal{G} : u_0(x) \mapsto u(x,t).
\label{eq:burgers_operator}
\end{equation}
Initial conditions are sampled from a Gaussian random field and reference
solutions are obtained from the released data generator for this benchmark,
which integrates \cref{eq:burgers} numerically. The trunk is evaluated on a
$50 \times 50$ grid covering $x$ and $t$, and both arms are trained on identical
data, so the comparison isolates the trunk embedding.

\begin{figure}[H]
\centering
\includegraphics[width=\textwidth]{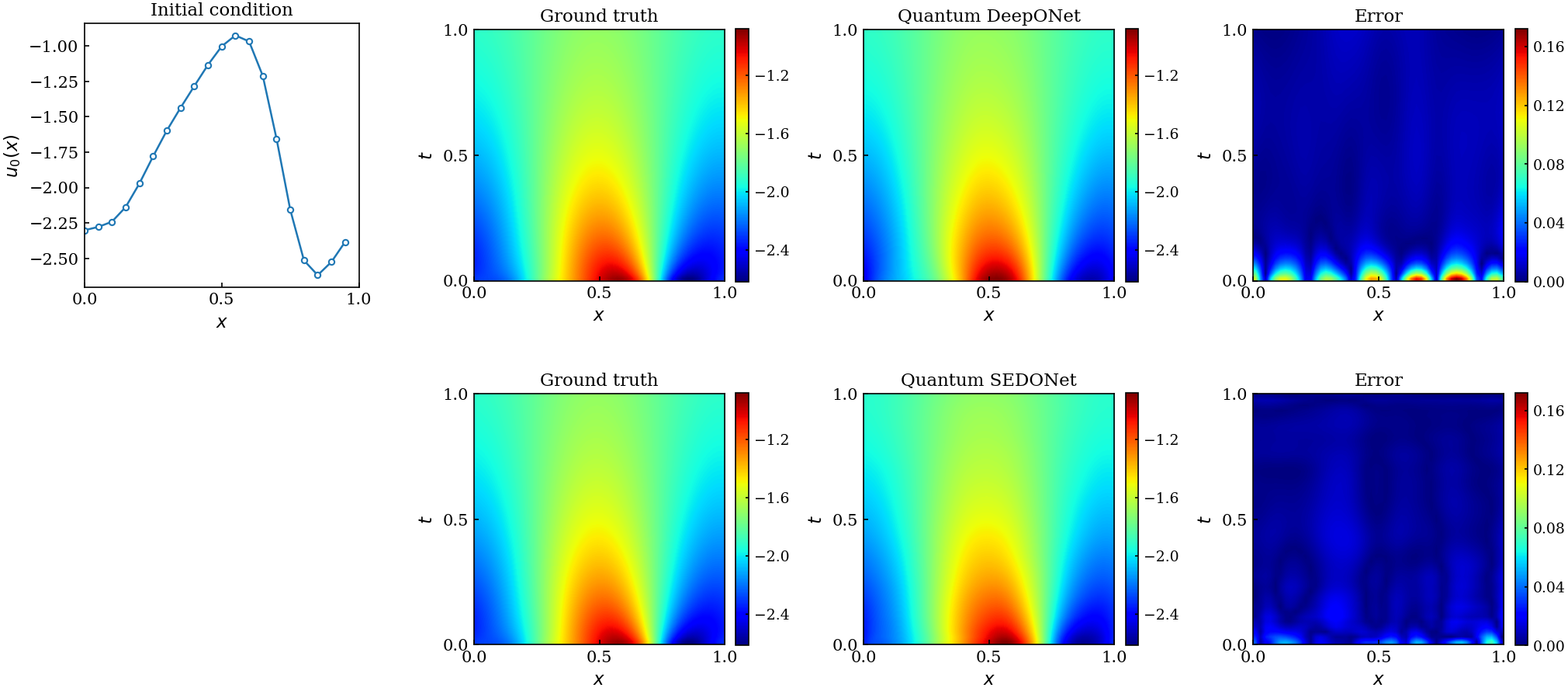}
\caption{Prediction comparison for one representative Burgers equation test
sample, showing the initial condition, the ground truth, Quantum DeepONet and
Quantum SEDONet predictions, and their corresponding absolute error maps on a
common color scale.}
\label{fig:burgers_fields}
\end{figure}

\Cref{fig:burgers_fields} shows a representative test function for the Burgers
benchmark. Structure is smoothed away rapidly and the difficulty is compressed
into the initial instants: the baseline error (top right) forms a bright band
along $t=0$, where the initial condition still carries the sharp features of the
Gaussian random field and the raw temporal coordinate leaves the trunk least
able to resolve them. Quantum SEDONet (bottom right) suppresses this band
substantially, its Chebyshev polynomial features providing a boundary-adapted representation the
boundary layer demands. The measured error concentration, the ratio of mean
error at $t=0$ to that at $t>0.2$, falls from roughly twenty to roughly ten,
confirming that the improvement acts on the predicted failure mode rather than
uniformly.

\begin{figure}[H]
\centering
\includegraphics[width=0.5\textwidth]{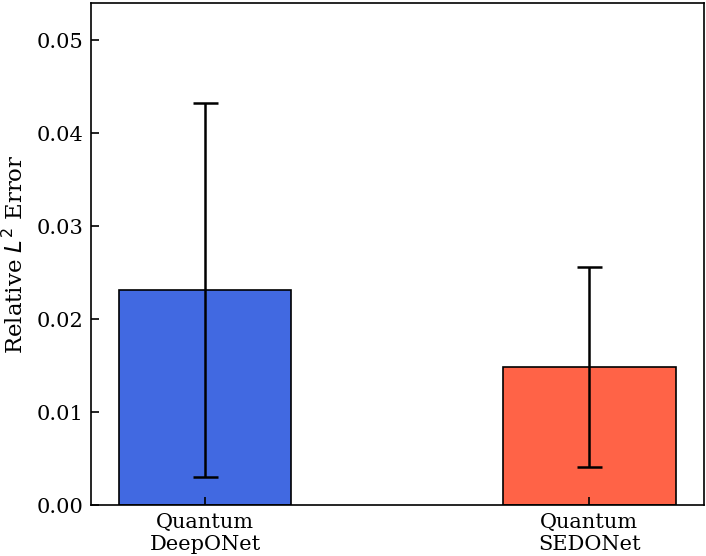}
\caption{Burgers: mean relative $L^2$ error over $100$ test functions, with
standard-deviation bars. Quantum SEDONet improves on the baseline for $93$ of
$100$ functions (paired $t=8.1$).}
\label{fig:burgers_bar}
\end{figure}

\Cref{fig:burgers_bar} shows the aggregate comparison over all $100$ Burgers
test functions: the mean relative $L^2$ error falls from $2.313\%$ to $1.481\%$,
a $36.0\%$ reduction. Paired, Quantum SEDONet improves on the baseline for $93$
of $100$ functions, with a paired $t$-statistic of $8.1$. The smaller
improvement relative to advection is consistent with the mechanism: the Burgers
solution is smoothed by viscosity over most of the domain, so the trunk faces
sharp structure only near $t=0$, whereas advection transports fine detail across
the whole space--time domain and its target operator is a pure translation that
Fourier features represent exactly.

\subsection{Channel Poisson Equation}
\label{sec:channel}
The advection and Burgers benchmarks are both periodic in space and non-periodic
in time, so the Chebyshev basis is always assigned to the temporal coordinate,
while the antiderivative has no periodic coordinate at all. To test whether the
boundary-matching principle holds when both spatial coordinates carry
boundary conditions, we add a channel Poisson problem,
\begin{equation}
-(u_{xx}+u_{yy}) = f(x,y), \qquad (x,y)\in[0,1]^2,
\label{eq:channel}
\end{equation}
periodic in $x$ ($u(0,y)=u(1,y)$) and Dirichlet in $y$ ($u(x,0)=u(x,1)=0$). This
is the first benchmark in which the two spectral bases are chosen from purely
spatial boundary conditions on a single problem: Fourier for the periodic
coordinate $x$ and Chebyshev for the walled coordinate $y$. The source fields are
Gaussian random fields that are periodic in $x$ and vanish at the walls, and
reference solutions are computed exactly in $x$ (by fast Fourier transform,
which decouples the problem into independent one-dimensional modes) and to
second order in $y$ (by a tridiagonal solve), with the Dirichlet condition
satisfied to machine precision. 

\begin{figure}[H]
\centering
\includegraphics[width=\textwidth]{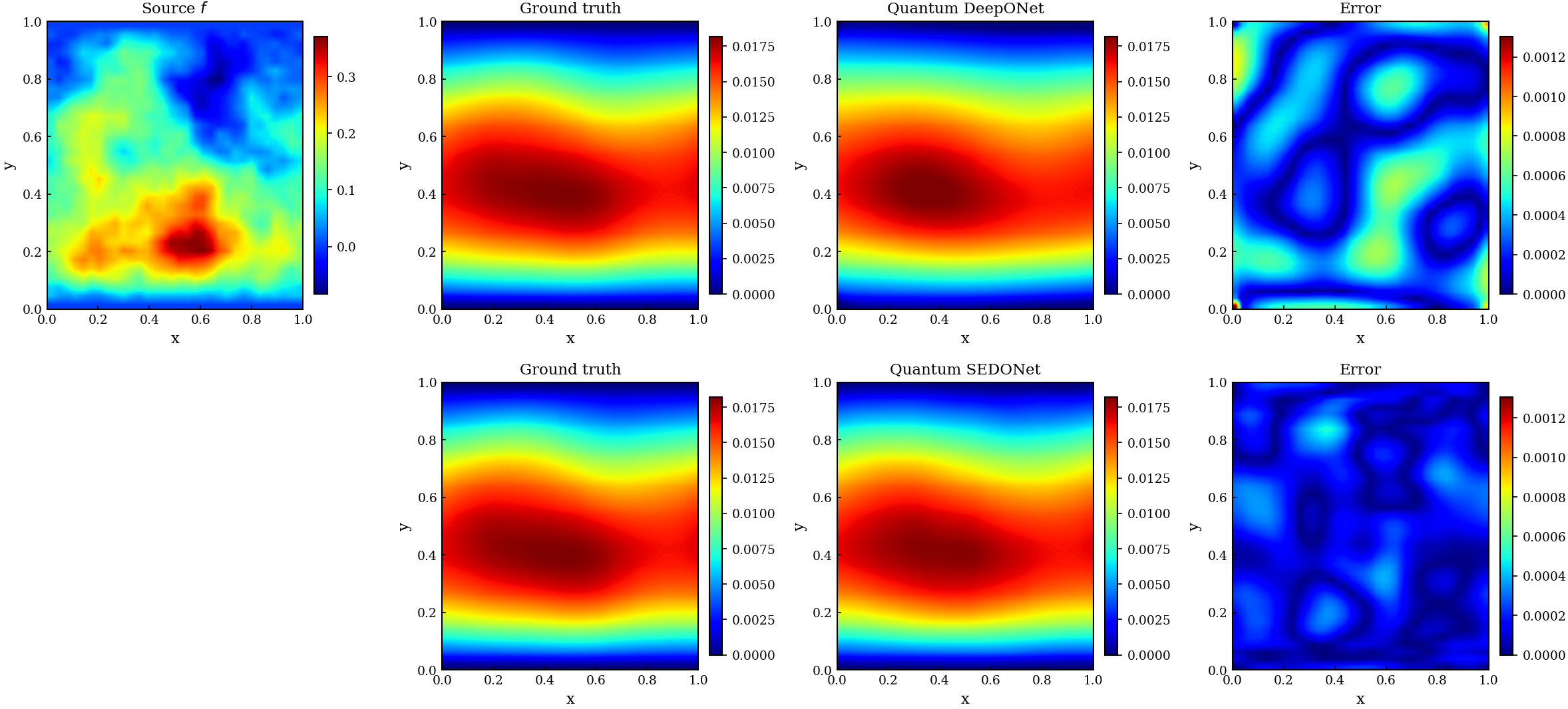}
\caption{Prediction comparison for one representative channel Poisson test
sample, showing the source field, the ground truth, Quantum DeepONet and Quantum
SEDONet predictions, and their corresponding absolute error maps on a common
color scale. The solution is periodic in $x$ and satisfies homogeneous Dirichlet
conditions at $y=0,1$. Quantum SEDONet reduces the error where the wall-normal
structure is sharpest.}
\label{fig:channel_fields}
\end{figure}

\Cref{fig:channel_fields} shows a representative test sample. The solution is
smooth, Poisson is a doubly-integrating, strongly smoothing operator and is
banded in $y$, pinned to zero at both walls, and modulated gently in the periodic
direction. The baseline (top row) leaves a structured error wherever the
solution has its sharpest wall-normal variation, which the raw $(x,y)$ trunk
resolves poorly; Quantum SEDONet (bottom row) suppresses this error
substantially, its Fourier terms matching the periodic $x$-structure and its
Chebyshev nodes resolving the wall-normal $y$-profile.

\begin{figure}[H]
\centering
\includegraphics[width=0.5\textwidth]{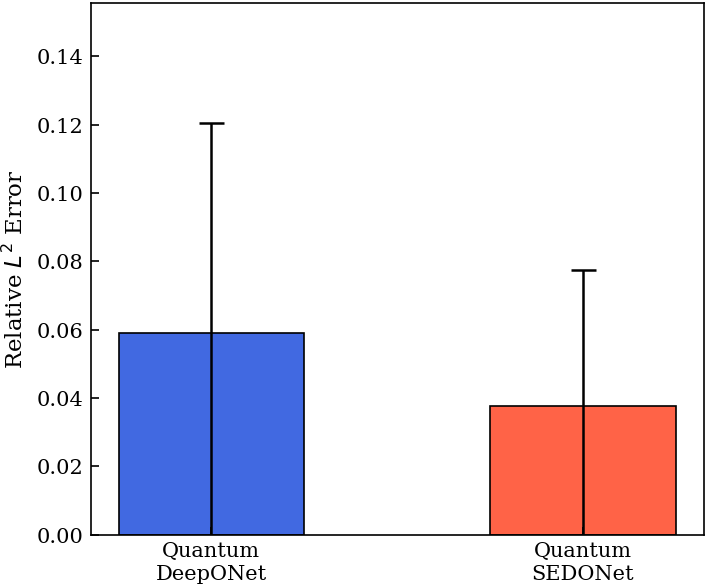}
\caption{Channel Poisson: mean relative $L^2$ error over the test functions,
with standard-deviation bars. Quantum SEDONet improves on the baseline for
$98\%$ of functions (paired $t=26.8$).}
\label{fig:channel_bar}
\end{figure}

\Cref{fig:channel_bar} shows the aggregate comparison over the test set: the
mean relative $L^2$ error falls from $5.889\%$ to $3.760\%$, a $36.2\%$
reduction. Paired, Quantum SEDONet improves on the baseline for $98\%$ of test
functions, with a paired $t$-statistic of $26.8$. The larger absolute errors
relative to the other benchmarks reflect the principal-component reduction of
the source, which caps the accuracy of both arms equally; the improvement itself
is comparable in magnitude to the Burgers gain, consistent with a problem whose
smooth solution offers no phase-rotation bonus of the kind that made advection
exceptional.

\subsection{The mechanism is consistent across all four problems}
\label{sec:mechanism}
The four benchmarks are physically distinct, one an elementary integration, one
convective, one diffusion-dominated, and one elliptic, yet the improvement tracks
the same cause in each. On the antiderivative the gain comes from replacing a
single affine coordinate, which a ReLU trunk can only extend into a
piecewise-linear approximation, with a polynomial basis matched to the bounded
non-periodic domain; the width-matched control shows that additional capacity
alone does not reproduce it. On Advection the gain comes from matching a
pure-translation operator with a Fourier basis, which represents its action as
an exact phase rotation, while the Chebyshev embedding supplies the temporal
resolution the raw coordinate lacked. On Burgers it comes from resolving the
$t=0$ boundary layer that a raw temporal coordinate cannot. On channel Poisson,
it comes from matching the periodic $x$-structure with Fourier features and the
wall-normal $y$-profile with Chebyshev nodes, the first case in which both
bases are dictated by spatial boundary conditions.

That a single principle, assigning each coordinate the spectral basis its boundary
condition dictates, produces the predicted improvement across integrative,
convective, diffusive, and elliptic regimes and across Chebyshev-only,
periodic-space with non-periodic-time, and mixed Dirichlet-periodic spatial
coordinates, is the central evidence that Quantum SEDONet addresses a real
representational deficiency rather than exploiting a quirk of one dataset. The
ordering of the gains, largest where the baseline representation is poorest
(the antiderivative's single raw coordinate) or where the operator has
exploitable structure (advection's exact phase rotation), and otherwise
comparable across problems, is itself consistent with the mechanism.

\subsection{Summary of results}
\label{sec:discussion}
Across the four benchmarks Quantum SEDONet reduces the mean relative $L^2$ error
by $54.1\%$ on the antiderivative operator, $49.6\%$ on advection, $36.0\%$ on
Burgers, and $36.2\%$ on channel Poisson, improving on the baseline for the
overwhelming majority of test functions in every case under a paired comparison.
In each instance the qubit count and circuit depth are identical to the
baseline, and the quantum and classical evaluation paths agree to within the
precision of the classical arithmetic, so every figure holds unchanged for ideal
quantum inference. \Cref{tab:main,tab:cost} in \cref{sec:summary} collect the
accuracy and resource comparisons.

The result is a rare case in which a representational upgrade is free in the
resource that constrains the platform. On classical hardware a richer input
embedding costs a proportional increase in the width of the first layer and
hence in compute. Under unary encoding the qubit count is set by the maximum of
the input and output dimensions of a layer, and the network width already
saturates that maximum; an embedding that stays at or below the width therefore
rides for free. The width-matched control on the antiderivative sharpens the
point: spending four extra qubits on trunk width made the fit worse, while
spending nothing and changing the basis more than halved the error. Three limitations bound these claims: all results are in ideal, noiseless
simulation; each arm is trained from a single seed, so the paired statistics
characterize variability across inputs rather than initializations; and all four
benchmarks have smooth solutions, the regime most favorable to global spectral
bases. \Cref{sec:future} takes up each of these.
\section{Summary and Conclusions}
\label{sec:summary}

We introduced Quantum SEDONet, a spectral trunk embedding for quantum deep
operator networks that assigns each query coordinate the classical spectral
basis its boundary condition dictates: a Fourier expansion for periodic
coordinates and a Chebyshev expansion for bounded, non-periodic ones. The method
leaves the Quantum DeepONet architecture, the orthogonal pyramidal layer, and
the tomography procedure entirely unchanged, and modifies only the classical map
applied to the trunk input before the first quantum layer. Under unary amplitude
encoding a layer's qubit count is one plus the larger of its input and output
dimensions, and the network width already saturates that maximum; an embedding
whose dimension stays at or below the width therefore costs no additional qubits
and no additional circuit depth.

\Cref{tab:main} collects the accuracy comparison. Across four benchmarks
spanning integrative, convective, diffusion-dominated, and elliptic regimes, the
embedding reduces the mean relative $L^2$ error by between $36\%$ and $54.1\%$,
and under a paired comparison on identical test functions it improves on the
baseline for the overwhelming majority of inputs on every problem. The
antiderivative operator, discussed in \cref{sec:antiderivative}, is treated
separately in the text: its trunk has a single bounded non-periodic coordinate,
so its embedding is Chebyshev alone, and it records the largest reduction of the
four at $54.1\%$ ($1.159\%\to0.532\%$, better on $99$ of $100$ test functions).

\begin{table}[H]
\centering
\caption{Mean relative $L^2$ error, baseline versus Quantum SEDONet.
``Better on'' counts test functions on which Quantum SEDONet has the lower
error under a paired comparison on identical inputs; the corresponding paired
$t$-statistics are reported in \cref{sec:advection,sec:burgers,sec:channel}.
Results for the antiderivative operator, whose embedding is Chebyshev alone, are
given in \cref{sec:antiderivative}.}
\label{tab:main}
\begin{tabular}{lcccc}
\toprule
Benchmark & Q-DeepONet & Q-SEDONet & Reduction & Better on \\
\midrule
Antiderivative       & $1.159\%$ & $0.532\%$ & $54.1\%$ & $99/100$ \\
Advection       & $3.463\%$ & $1.744\%$ & $49.6\%$ & $199/200$ \\
Burgers         & $2.313\%$ & $1.481\%$ & $36.0\%$ & $93/100$  \\
Channel Poisson & $5.889\%$ & $3.760\%$ & $36.2\%$ & $98/100$  \\
\bottomrule
\end{tabular}
\end{table}

\Cref{tab:cost} records the resource accounting that makes the method
attractive on near-term hardware. On every benchmark the qubit count and the
circuit depth are unchanged, and the number of evaluated circuits is identical;
the sole overhead is a few percent more trainable parameters, contributed by the
additional RBS angles in the first trunk layer. The antiderivative case,
omitted from the table, follows the same pattern with an unchanged trunk
register of $11$ qubits.

\begin{table}[H]
\centering
\caption{Resource cost of the spectral trunk embedding. Qubit count and circuit
depth are unchanged on every benchmark. The antiderivative operator, whose trunk register
is likewise unchanged at $11$ qubits, is described in
\cref{sec:antiderivative}.}
\label{tab:cost}
\begin{tabular}{lcccc}
\toprule
Benchmark & Baseline trunk & SEDONet trunk & Qubits & Parameters \\
\midrule
Advection       & raw $(x,t)$        & Fourier$(x)+$Cheb$(t)$ & $22\to22$ & $3544\to3669$ ($+3.5\%$) \\
Burgers         & raw $(x,t)$      & Fourier$(x)+$Cheb$(t)$ & $21\to21$ & $2870\to2940$ ($+2.4\%$) \\
Channel Poisson & raw $(x,y)$        & Fourier$(x)+$Cheb$(y)$ & $21\to21$ & $2769\to2884$ ($+4.2\%$) \\
\bottomrule
\end{tabular}
\end{table}

The improvement tracks a diagnosed failure mode rather than appearing
uniformly. Where a coordinate reaches the trunk as a single affine input, the
network must assemble every oscillation in that direction from its
nonlinearities, and the resulting error concentrates wherever the solution
varies fastest: along the transport characteristics for advection, in the $t=0$
boundary layer for Burgers, at the sharpest wall-normal variation for channel
Poisson, and as visible piecewise-linear kinking for the antiderivative. In each
case supplying the matching spectral basis removes precisely that structure.
Alongside the operator benchmarks we reproduced the function-approximation tests
of the original Quantum DeepONet study to validate the classical-to-quantum
parameter transfer end to end, and we identified and corrected a numerical
instability in the reference unary data loader that otherwise prevents the
simulation of networks whose activations contain exact zeros.
 
\section{Future Work}
\label{sec:future}
Two steps are prerequisites for stronger claims rather than extensions of them.
All results reported here are obtained in the ideal, noiseless, infinite-shot
regime; repeating the comparison under a depolarizing channel and a
device-calibrated backend noise model, with finite shot counts and unary
post-selection during tomography, would let the accuracy advantage be stated
under realistic hardware conditions. Because Quantum SEDONet has exactly the
same circuit depth and qubit count as the baseline, we expect that advantage to
survive a fixed noise model, but the outcome is not obvious: the induced error
depends on the RBS angle values themselves, and the two arms converge to
different angles. Separately, every figure reported here comes from a single
training seed per arm. The paired statistics characterize variability across
test inputs, not across initializations, and these are distinct sources of
uncertainty; a multi-seed study reporting mean and standard deviation per arm is
the primary outstanding validation step.

Three directions extend the method itself. First, the number of Fourier and
Chebyshev terms was fixed by a resource-driven rule, expanding each coordinate
until the embedding fills the network width, rather than an accuracy-driven one.
The optimal degree is likely problem-dependent, since a solution with fine
structure in one direction and smooth variation in another would be better
served by an uneven split; selecting the degree adaptively is a natural
refinement, and it becomes necessary in higher-dimensional and multiphysics
systems, where more coordinates compete for the same width budget. Second, the
rule as stated covers periodic and bounded non-periodic coordinates. Extending
it to Neumann conditions, to semi-infinite and unbounded domains, where Laguerre
and Hermite expansions are the classical choice, and to geometries that do not
factor coordinate-wise would establish how far the boundary-matching principle
carries. Third, every benchmark considered here has a smooth or strongly
smoothed solution, precisely the regime in which global polynomial and
trigonometric bases are most effective. A problem with genuine interior
discontinuities, such as a shock or a material interface, would test the method
where global spectral bases classically struggle; we would expect the margin to
narrow there, and localized or multi-domain bases may be the appropriate
response.

\section*{Declaration of Competing Interest}
The authors declare that they have no known competing financial interests or personal relationships that could have appeared to influence the work reported in this paper.

\section*{Acknowledgements}
This work was supported in part by the 

\section*{Data Availability}
The data supporting the findings of this study are available from the corresponding author upon reasonable request.


\bibliographystyle{unsrt}
\bibliography{references}

\end{document}